\documentclass{JINST}
\usepackage{graphicx}
\usepackage[figuresright]{rotating}

\RequirePackage{lineno}
\title{Liquid Argon Time Projection Chamber Research and Development in the United States}         

\author{B.~Baller$^a$,
C.~Bromberg$^b$,
N.~Buchanan$^c$,
F.~Cavanna$^d$,
H.~Chen$^e$,
E.~Church$^d$,
V.~Gehman$^f$,
H.~Greenlee$^a$,
E.~Guardincerri$^g$,
B.~Jones $^h$,
T.~Junk$^a$,
T.~Katori$^h$,
M.~Kirby$^a$,
K.~Lang$^i$,
B.~Loer$^a$,
A.~Marchionni$^a$,
T.~Maruyama$^j$,
C.~Mauger$^g$,
A.~Menegolli$^k$,
D.~Montanari$^a$,
S.~Mufson$^l$,
B.~Norris$^a$,
S.~Pordes$^a$,
J.~Raaf$^a$,
B.~Rebel$^a$,
R.~Sanders$^a$,
M.~Soderberg$^{a,m}$, 
J.~St.~John$^n$,
T.~Strauss$^o$,
A.~Szelc$^d$,
C.~Touramanis$^p$,
C.~Thorn$^e$,
J.~Urheim$^l$,
R.~Van de Water$^g$,
H.~Wang$^q$,
B.~Yu$^e$,
M.~Zuckerbrot$^a$\\
\llap{$^a$}Fermi National Accelerator Laboratory, Batavia, IL 60510, USA\\
\llap{$^b$}Michigan State University, East Lansing, MI 48824, USA\\
\llap{$^c$}Colorado State University, Fort Collins, CO 80523, USA\\
\llap{$^d$}Yale University, New Haven, CT 06520, USA\\
\llap{$^e$}Brookhaven National Laboratory, Upton, NY 11973, USA\\
\llap{$^f$}Lawrence Berkeley National Laboratory, Berkeley, CA 94720, USA\\
\llap{$^g$}Los Alamos National Laboratory, Los Alamos, NM 87545, USA\\
\llap{$^h$}Massachusetts Institute of Technology, Cambridge, MA 02139, USA\\
\llap{$^i$}University of Texas at Austin, TX 78712, USA\\
\llap{$^j$}High Energy Accelerator Research Organization, KEK, Tsukuba, Ibaraki 305-0801, Japan\\
\llap{$^k$}Istituto Nazionale di Fisica Nucleare, Pavia 6-27100, Italy\\
\llap{$^l$}Indiana University, Bloomington, IN 47405, USA\\
\llap{$^m$}Syracuse University, NY 13210, USA\\
\llap{$^n$}University of Cincinnati, Cincinnati, OH 45220, USA\\
\llap{$^o$}University of Bern, 3012 Bern, Switzerland\\
\llap{$^p$}University of Liverpool, Liverpool, Merseyside L69 3BX, United Kingdom\\
\llap{$^q$}University of California Los Angeles, Los Angeles, CA 90095, USA}

\abstract{A workshop was held at Fermilab on March 20-21, 2013 to discuss the development of liquid argon time projection chambers (LArTPCs) in the United States.  The workshop was organized under the auspices of the Coordinating Panel for Advanced Detectors, a body that was initiated by the American Physical Society Division of Particles and Fields.  All presentations at the workshop were made in seven topical plenary sessions:  $i)$ Argon Purity, $ii)$ Cryogenics, $iii)$ TPC and High Voltage, $iv)$ Electronics, Data Acquisition and Triggering, $v)$ Scintillation Light Detection, $vi)$ Calibration and Test Beams, and $vii)$ Software.  This document summarizes the current efforts in each of these areas.  It also highlights areas in LArTPC research and development that are common between neutrino experiments and dark matter experiments.
}


\begin{document}

\section{Introduction}
\label{sec:Introduction}
Liquid argon time projection chambers (LArTPCs) are a promising technology for use in both neutrino experiments and dark matter experiments.  A  minimum ionizing charged particle passing through the liquid argon produces 55,000 electrons for every centimeter traversed.  Those electrons are drifted in a uniform electric field toward a readout that records the current created by the drifting electrons.  Because the electric field is uniform, the drift of the electrons is a constant velocity, allowing one to determine the location in the drift direction at which the ionization occurred with exceptional resolution.  This resolution provides neutrino experiments the ability to digitally record bubble-chamber-like images of the neutrino interactions, allowing researchers to distinguish between different interaction processes with certainty.  Because the liquid argon also scintillates as charged particles pass through it, emitting 80,000 photons for every centimeter traversed by a minimum ionizing particle, it can be used for dark matter detection experiments as well.  In a dark matter experiment both the light and charge information are used to distinguish candidate interactions from background.

The ICARUS~\cite{Amerio:2004ze, Rubbia:2011ft} experiment was the first LArTPC having a mass of $> 100$ tons to be constructed in the world.  It has an active mass of over 500 tons of liquid argon, is located in the Gran Sasso Laboratory in Italy, and observes neutrinos from the CNGS beam.  It is the culmination of decades of R\&D efforts in Europe and the US.  It demonstrated that a LArTPC can be effectively used for neutrino physics by observing the CNGS neutrino beam.  However, a hundred ton scale LArTPC does not sufficient mass to answer the current experimental questions about neutrino oscillations, namely do neutrinos violate the CP symmetry and what is the hierarchy of the neutrino mass states.  Many efforts in the US have started in the last 10 years to resolve technical issues related to building a multi-kiloton scale LArTPC.  

This document is the summary of a workshop held at Fermilab March 20-21, 2013~\cite{workshop} to discuss the current efforts to develop LArTPC detectors in the US.  The workshop was organized under the auspices of the Coordinating Panel for Advanced Detectors (CPAD), a standing panel that was empowered by the American Physical Society Division of Particles and Fields (DPF).  The DPF convened a working group to determine the best way to keep the development of particle detector technology vibrant in the US.  That working group proposed the organization of a standing panel intended to stimulate the research and development of these detectors; that panel is CPAD.  It is intended to promote national detector R\&D programs and stimulate new ideas in instrumentation development.  CPAD also facilitates coordination among the national HEP laboratories and university groups engaged in detector R\&D and the use of targeted resources at the national laboratories.  It is not to act as an advisory committee, a standing review body for proposals or to provide a set direction for the national detector R\&D program.  The CPAD group is also involved in the 2013 Community Summer Study~\cite{snowmass} through the Instrumentation Frontier.  

The CPAD group has organized several working groups, among which is the detector system development group.  The conveners of that working group, David Lissauer from Brookhaven National Laboratory and Ed Blucher from the University of Chicago organized a topical group on LArTPCs to be convened by Brian Rebel of Fermilab, Craig Thorn of Brookhaven National Lab and Jon Urheim of Indiana University.  

Rebel, Thorn and Urheim organized the Fermilab workshop, which was attended by 72 individuals from 31 institutions.  The workshop primarily focused on LArTPCs applied to neutrino oscillation physics but synergies with dark matter efforts were discussed as well.  All presentations at the workshop were made in plenary sessions organized into seven topical categories:  $i)$ Argon Purity, $ii)$ Cryogenics, $iii)$ TPC and High Voltage, $iv)$ Electronics, Data Acquisition and Triggering, $v)$ Scintillation Light Detection, $vi)$ Calibration and Test Beams, and $vii)$ Software.  This document summaries the information presented in those sessions in the sections below.  The work for several efforts throughout the US can be placed in more than one of the topical categories.  As such, the reader will find aspects of those projects described throughout the document.  Important lessons learned in any of those areas that are key to building LArTPC experiments are preceded by a "{\bf Learned Lesson}" tag in the text. Each section describes which groups are participating in the topical category, synergies between those efforts and with dark matter experiments, and outstanding issues that need to be addressed before the construction of a multi-kiloton scale LArTPC, like the LBNE far detector, can begin.

\section{Physics Requirements for Neutrino and Dark Matter Experiments}
\label{sec:PhysicsRequirements}
\subsection{Neutrino Experiments}

The major physics measurements proposed for large LArTPCs in long baseline experiments are the complete and precise determination of the mass hierarchy as well as the parameters defining the PMNS mixing matrix, particularly the CP violating phase $\delta$.  The detection of a supernova core-collapse event, relic neutrinos, and the determination a larger least upper bound for the proton partial lifetime in one or more important decay modes are also major goals that can be accomplished with very large LArTPCs located deep underground.  Other objectives of current interest requiring smaller LArTPCs are the resolution of the conflicting observations of short baseline neutrino oscillations, with possible implications for the observation of sterile neutrinos, and the observation of coherent neutrino scattering and measurements of cross sections for this process.  Successfully achieving these goals with LArTPCs will require additional studies to develop a precise understanding of cross sections for neutrino and anti neutrino interactions in argon, and a detailed knowledge of the final state interactions and other nuclear effects.

The successful achievement of these physics goals places many demands on the performance of LArTPCs, and on our understanding of the processes involved in producing the observed charge and light signals and on the backgrounds that accompany the desired signals.  For the precise determination of cross sections, clean identification of the final reaction products is required.  This goal also demands precise energy loss and range determination, precise total energy measurement for isolated tracks as well as showers, and high spatial resolution, especially near the vertex.  In general, the high spatial resolution and ability to separate multiple closely spaced tracks gives LArTPC the ability to use topological classification to help identify events in the presence of final state interactions and to reconstruct total energy. To properly separate neutral and charged current interactions, good separation of electron gamma showers obtained by observing the charge density at the beginning of the showers is required.  The identification of final state interactions will require low energy thresholds in order to observe, for example, Compton scattered photons from nuclear de-excitation and low energy argon recoils from neutrons.  To separate neutrino and anti-neutrino charge current interactions, charge determination of the outgoing muon is required.  This separation can be done by creating a magnetic field in the detector, or alternatively by observing $\mu^{-}$ capture.  In the latter case, the capture cross section needs to be well determined.  For supernova neutrino detection and coherent neutrino-nucleus scattering low energy detection thresholds will be required.  For coherent neutrino scattering, in particular, the thresholds are so low (keV) that scintillation light is the only useful signal available.

Many research and development programs are presently being pursued to improve the performance of the next generation of LArTPCs to meet the physics requirements of future experiments.  Examples of studies in progress to define device parameters are
\begin{itemize}
\item	Selection of wire geometry to optimize signal formation, stereo reconstruction, and to minimize noise
\item	Selection of wire pitch to optimize particle identification by energy loss as a function of range and electron from photon separation
\item	Operation of electronic systems in liquid argon
\item	Impact of diffusion of electrons and Rayleigh scattering of scintillation light in very large detectors
\item	Measurement of electron-ion recombination, especially at high specific energy loss, to allow accurate conversion of observed charge into deposited energy
\item	Measurement of electron attachment to impurities
\item	Understanding  generation,  transport, and ultimate limits of impurities in LArTPCs
\item	Understanding mechanisms and limits of high voltage breakdown in liquid argon
\item	Reducing noise sources for charge detection, and increasing system quantum  efficiency for photon detection, to decrease the energy threshold for low energy events such as Compton electrons and nuclear recoils
\item	Consequences of high cosmic muon background in surface detectors, including track distortions produced by space charge due to positive ions
\item	Reducing backgrounds from radioactive contaminants in the liquid argon
\item	Understanding and reducing backgrounds from muon interactions in the shielding rock that produce neutrals that enter the detector and charge exchange
\end{itemize}

\subsection{Dark Matter Experiments}

For most experimental dark matter searches the detection candidate is a weakly interacting massive particle (WIMP), a particle that acts like heavy neutrino with only neutral-current interactions.   Events are therefore ultra low in rate, less than $10^{-3}$/day/kg and very low in energy, between 10-100 keV, making background reduction and rejection the overriding experimental consideration.  Since, at these low energies the events are point-like, good spatial resolution is not required, other than for background rejection.  Typically these detectors are dual phase electroluminescence TPCs, achieving separation of electromagnetic and nuclear recoil interactions by observing scintillation light from the primary ionization event in liquid argon, which is the S1 scintillation signal, and from electroluminescence created by accelerating the extracted ionization electrons in the gas space above the liquid, which is the S2 scintillation signal.  Nuclear recoils have high $dE/dx$ compared to electrons, and therefore higher ion recombination and higher singlet excitation.  The result is that recoils have faster scintillation pulses and lower electron extraction to scintillation ratios (S2/S1) than electromagnetic interactions.  Since the magnitude and time distribution of light from the scintillations are the signals of interest for particle discrimination, and since the photon yield is low at the energies of interest, very high light collection is needed, namely more than 8 photo electrons/keV.  Low photon absorption at 128~nm, low electron attachment, efficient and uniform wavelength shifters and light reflectors, efficient and stable electron transport, and high quantum efficiency photdetectors are the principal requirements for good signal formation.

Using pulse shape discrimination suppresses the background from $^{39}Ar$, leaving the residual neutron background which
comes from two sources: $i$) radiogenic , such as U and Th in the detector materials and the experiment hall, and $ii)$ cosmogenic, produced by cosmic ray showers, with energies up to GeV, and therefore very penetrating.  The radiogenic neutrons are a few MeV in energy and since they are not very penetrating, they can be reduced by very clean detector materials, passive shielding, and active vetoes.  Cosmogenic neutrons are reduced by locating experiments deep underground and vetoing the muons and showers.  The combined consequence is that detectors must be made of ultra-clean materials in clean room environment and sited deep underground with efficient active shielding.  Detector radiologic cleanliness required the use of Ar depleted in $^{39}$Ar that is chemically purified from underground sources, highly purified construction materials, and detector construction in high quality clean-room environments.

\section{Argon Purity}
\label{sec:Purity}
Fermilab has led the US efforts at understanding how to keep LArTPCs free from electronegative contamination.  It has assembled two test stands for that purpose, the Materials Test Stand (MTS) and the Liquid Argon Purity Demonstrator (LAPD).  The amount of electronegative contamination, or impurity, in the liquid argon determines the electron drift lifetime.  That is, the more impure the argon, the lower the drift lifetime and the smaller the drift distance must be in order to see an ionization signal from the readout.

\subsection{The Materials Test Stand}
\label{sec:MTSP}

\begin{figure}
\centering
\includegraphics[width=3.in]{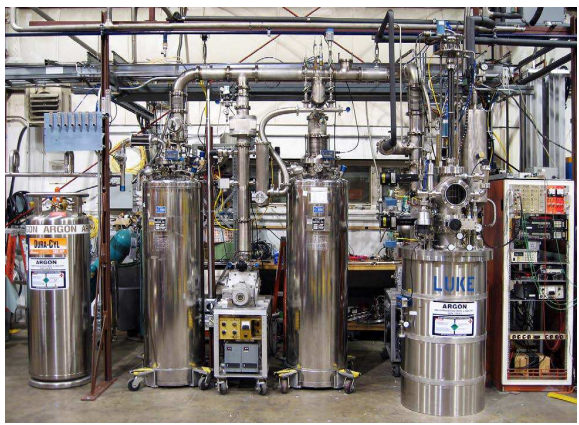}
\caption{Photograph of the materials test system (MTS) at FNAL. Commercial argon (left) is passed through two filters (center) before entering the cryostat (right).}
\label{fig:mtsphoto}
\end{figure}

\begin{figure}
\centering
\includegraphics[width=3.in]{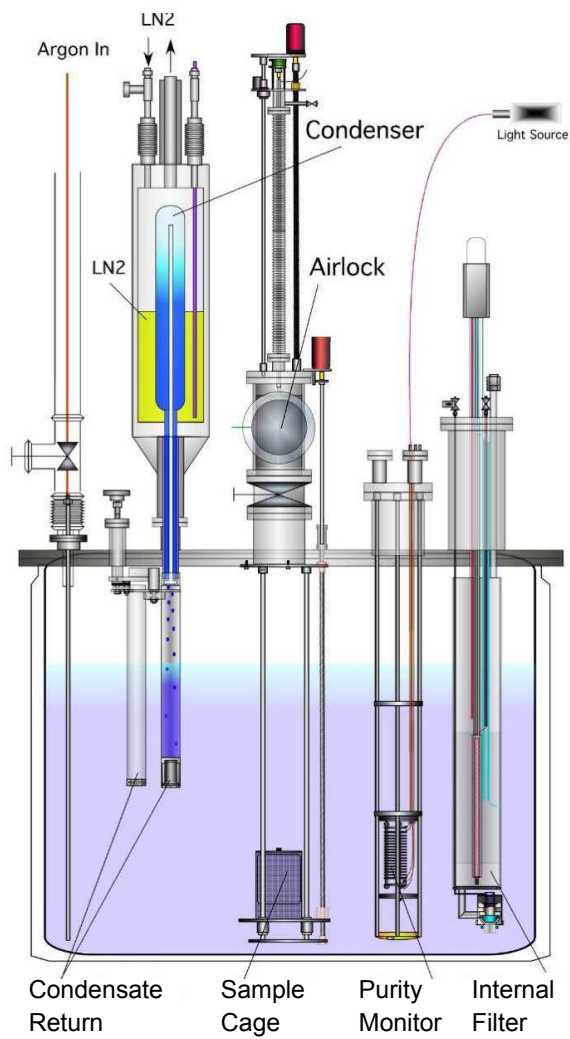}
\caption{The MTS cryostat.}
\label{fig:MTS}
\end{figure}

The MTS effort is led by Pordes, with Jaskierny, Jostlein,  Kendziora, Para, Skup, and Tope contributing as well.  The MTS is shown in Fig.~\ref{fig:mtsphoto} and is decribed in detail in~\cite{MTS}.  The supply argon is sent through a molecular sieve to remove water and then through activated copper to remove oxygen before entering the evacuated cryostat.  The cryostat has a capacity of 250~L and is vacuum insulated.  Figure~\ref{fig:MTS} shows a schematic of the cryostat and its instrumentation.  The heart of the MTS is the airlock-sample cage assembly located in the center of the schematic.  The airlock allows materials to be introduced into the cryostat without first having to empty it of liquid argon.  The sample cage can be positioned anywhere between the airlock and the bottom of the cryostat to understand the temperature dependence of the impurities introduced by different materials. 

In addition to the airlock, the cryostat also contains an ICARUS style purity monitor~\cite{icarus} to determine the electron lifetime and an internal filter to remove impurities introduced by the materials placed into the cryostat.  The cryostat can be operated as a closed system when the liquid nitrogen cooled condenser is in operation.  The argon gas that passes into the condenser is liquified and returned into the bulk liquid.  

It has been on operation for over 5 years and has tested a variety of materials for their suitability of use in LArTPCs.  The complete list can be found online~\cite{MTStable} and the primary results are described  in~\cite{MTS}.  The tests using the MTS have found no solid material which when it is immersed in liquid argon affected the drift lifetime. The problems come from out-gassing by material in the warm(er) regions of the argon vapor. The effect of a given material depends sensitively on its temperature and maybe even more sensitively on the pattern of the flow of argon vapor. If, for example, the flow of the vapor is away from the surface and at a rate to counteract any diffusion towards the liquid, the effect of the out-gassing will be minimal. As such the MTS is capable of comparing materials and giving repeated test to the statement that no material in the liquid has affected the drift-lifetime. The results reported by the MTS are essentially comparative numbers - water concentrations produced and effects on lifetime in various conditions. The facility is open to use by any group which cares to submit a sample for testing.

The MTS was an essential test bed for also understanding how to operate a closed system of liquid argon that would be pure enough to achieve an electron drift-lifetime of many milliseconds.   Adam Para was responsible for stimulating work towards massive liquid argon TPCs at Fermilab. Given the large body of knowledge and experience that already existed in Europe thanks to the ICARUS program, the Fermilab effort chose to develop a facility for testing candidate detector materials which could operate successfully in a cryostat which could and would not be evacuated. The motivation was that the cryostat for any multi-kiloton LArTPC would not be evacuable and the benefits of pumping on the cryostat and the detector inside which would reduce outgassing from detector materials would therefore not be available.

There were two challenges in achieving the ability to test materials: making clean argon and inserting materials into this argon.  The Fermilab group decided to make a system to clean the argon locally rather than use commercial filters for two reasons. Long term, it was believed this option would be cheaper and it would further the understanding of their operation, in particular the action of the water removal and the oxygen removal independently. The actual filter material is purchased from commercial vendors but the enclosures that house the material and the system that allows the filter material to be regenerated in place were designed at Fermilab.

The MTS first operated with an open system and achieved many millisecond electron drift lifetimes. After this goal was accomplished a condenser was added to the system to maintain the argon and control the MTS cryostat pressure in January 2008.  Figure~\ref{fig:MTS} shows the condenser as it is now implemented. The coolant is liquid nitrogen; argon gas goes up the outside annulus, is condensed by the nitrogen and the liquid flows down the inner pipe. The original implementation had nothing below the level of the cryostat flange and the condensate simply dripped back into the cryostat.

\begin{figure}
\centering
\includegraphics[width=3.in]{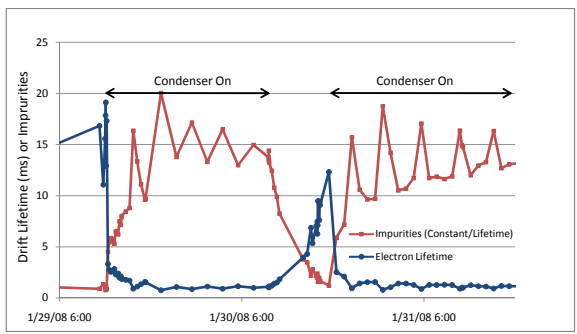}
\caption{Effect of condenser operation on electron drift lifetime.  The impurities, defined as a constant divided by the drift lifetime, represent the physical contaminants in the argon.  When the condenser is off, the drift lifetime approaches 20 ms; when the condenser is on, the lifetime quickly degrades to 1 ms or less.  The oscillations in the drift lifetime are related to cycling of the condenser.}
\label{condensereffect}
\end{figure}

It quickly became clear that condensing reduced the electron drift lifetime dramatically, from ten milliseconds to less than one millisecond, as shown in Fig.~\ref{condensereffect}. After several hours, the condenser was turned off and the lifetime recovered in a matter of hours.

This behavior was initially ascribed to faulty construction of the condenser, causing it to be dismantled and rebuilt, only to suffer the same fate again. The next hypothesis explored was that the argon became charged by its flow through the pipe.  To test this hypothesis a return pipe was installed into the liquid with steel wool at the end to discharge any ions. The steel wool produced a significant improvement, leading to the speculation that the condenser took some contaminant in the vapor and injected it efficiently into the liquid argon.  As described in~\cite{MTS}, particulate contaminants were ruled out and water from the warm surfaces of the cryostat above the liquid level was determined to be the contaminant.  Because water has an affinity for metal surfaces, the steel wool captured most of the water in the returned liquid. Similarly, the recovery of the drift lifetime once the condenser was turned off happens as the water injected into the liquid argon plates out on the metal surface of the cryostat.

\begin{figure}
\centering
\includegraphics[width=3.in]{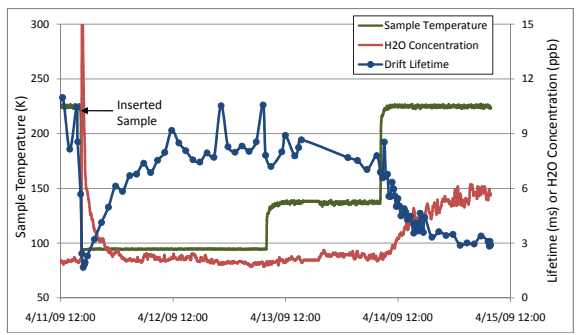}
\caption{Material test of {FR4} MicroBooNE y-plane wire endpoint.  The sample was first lowered into the liquid argon then raised into the vapor so that the temperature of the sample was increased.  When moved to 225 K, the sample began to outgas and the effect on water concentration and drift lifetime can be seen in the figure.  A similar relationship between water concentration and drift lifetime was observed during other material tests.}
\label{fr4materialtest}
\end{figure}

Once the water vapor injected into the liquid was suspected as the cause of the reduction in electron drift-lifetime, a water vapor monitor was acquired from Tiger Optics with ppb resolution.  The effect of various materials was measured with the monitor. The MTS design allows one to position the sample at any height inside the cryostat; the thermometer attached to the sample cage showed a strong temperature gradient above the liquid level.  Positioning the sample at various heights produced quite different amounts of out-gassing, which is consistent with the fact that the vapor pressure of water is strongly correlated with temperature. In general the product of the water concentration in the vapor and the electron drift-lifetime was a constant, independent of material, and causing water to be labeled as the main cause of poor drift lifetime. Figure~\ref{fr4materialtest} shows an example of this behavior.

{\bf Learned Lessons} The construction and operation of the MTS has provided several useful lessons: 
\begin{itemize}
\item While it is essential to test any candidate material in liquid argon, it is equally if not more important to measure its outgassing at temperatures which are elevated compared to 87K;
\item The temperature profile of the material will determine the amount it out-gasses; this fact applies to the metal surfaces of the cryostat which  can outgas water `for ever'; 
\item The pattern of the vapor flow will determine how much of this outgassing reaches the liquid directly;
\item The product of the two above effects is the most important determinant for the electron drift lifetime.
\end{itemize}

\subsection{LAPD}
\label{sec:LAPD}

\begin{figure}[!htbp]
\centering
\includegraphics[width=3in]{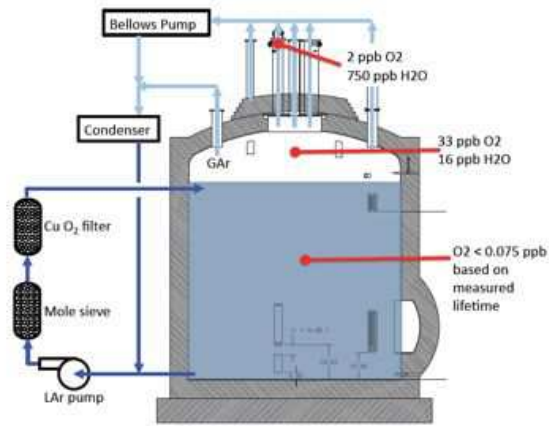}
\caption{ LAPD schematics and residual O2 and H2O concentration in gas and liquid argon phases. }
\label{fig:LAPD} 
\end{figure}

The Liquid Argon Purity Demonstrator (LAPD) is a system for liquid argon purity tests. This system was built by Fermilab, with Rebel, Plunkett and Tope leading the effort.  The LAPD tank can hold 22,000~L of liquid argon, but it cannot be evacuated.  Instead,  the tank air is removed by a purge of room temperature argon gas.  The effectiveness of  this method in  achieving very low concentration of residual air components was the main goal of LAPD. In addition to purging the interior of the vessel prior to filling with liquid argon, the LAPD system uses an 80 liter molecular sieve filter and 80 liter oxygen filter to remove contaminants from gaseous and liquid argon.  The system is shown schematically in Fig.\ref{fig:LAPD}. 

The first operation of the system was during Winter 2011-2012 and the inside of the tank was instrumented with 4 ICARUS style purity monitors.  An electron lifetime of over 3~ms was demonstrated during this run which validated the room temperature argon gas purge concept.  The second run started during Winter 2012-2013 and included a TPC with a 2 meter drift inside the tank.  Again, an electron lifetime in excess of 3~ms was achieved.  For comparison, the LBNE requirement on electron lifetime is 0.85~ms over a 2.32~m drift distance.  

The room temperature argon gas purge exchanged 7.5 volumes and achieved contamination levels below 6~ppm oxygen, 18~ppm nitrogen, and 1.2~ppm water in about 24 hours.  The room temperature argon gas in the tank was then re-circulated through  the filters for 77 volume changes which took about 1 week.  The oxygen dropped to 20~ppb and the water to about 0.6~ppm.  Nitrogen is not filtered but due to the mass ratio between argon vapor and argon liquid, 18~ppm nitrogen in the vapor will only add 21~ppb nitrogen to the argon liquid.  

The tank was filled through the filters with source liquid argon that contained 200~ppb oxygen and 8~ppm nitrogen.   After filling the liquid argon in the tank contained 30~ppb oxygen.  An electron lifetime of 1~ms was achieved after 6.6 volume changes of liquid recirculation through the filters and the volume exchange rate was 2.42 volumes changes per day.  The lifetime stabilized at an indicated 4~ms after 60 volume changes.  The pump flow rate was reduced from 2.45 to 1.22 and then from 1.22 down to 0.69 volume exchanges per day.  The electron lifetime remained constant despite the reduced filtration rate indicating that the boil off gas intercepts the contamination outgassing from the cables and other warm surfaces in the ullage space before it can diffusion into the liquid argon.  The boil off gas is condensed by a liquid nitrogen powered condenser which then feeds the condensate to the liquid pump suction where it is sent to the filters before returning to the tank.  The contamination in the ullage argon gas in the vicinity of the TPC signal feed through is quite high at 600~ppb oxygen and 2.4~ppm water.  During the initial 4 months of operation with the full volume of 22,000~L these filters have not saturated.  

{\bf Learned Lesson} A key lesson learned during the initial operation of LAPD is that virgin oxygen filter material must be very carefully regenerated to avoid damaging self heating.  The regeneration process involves slowly increasing the hydrogen fraction of the regeneration gas from 0\% to 0.35\%.  At 0.35\% hydrogen the filter exhibited 10~K/hr self heating.  During subsequent operation, a 4~inch diameter aperture in the tank vapor space was opened to remove a component from the tank and this removal did not introduce significant contamination into the tank.  

Entry into the tank between the operating periods revealed a surprising amount of particulate matter inside of the tank despite a 15 micron filter tank return filter.  

Another important LAPD goal was the study of temperature gradients in both liquid and gas phase.  Careful measurements of the bulk liquid temperature reveal it to be isothermal within 0.1 K except within a few inches of the liquid vapor interface where evaporation takes place.  Measurements of the temperature gradient in the vapor space in a half full tank show large temperature gradients on the order of 60 K over 20 inches of vapor immediately above the liquid.  

The LAPD system will continue operating through Summer 2013.  It will measure the capacity of the filter media for removing water.  It is also taking data using a 2~m drift distance LArTPC exposed to cosmic rays.

\section{Cryogenic Systems}
\label{sec:Cryogenics}
The experience gained through the design and construction of the LAPD system has been shared with other efforts throughout the US.  MicroBooNE, LBNE, LArIAT, and CAPTAIN are all reusing or improving designs initially developed with LAPD. Some of the primary lessons learned from the LAPD were how to efficiently construct the systems by using pre-insulated piping rather than vacuum-jacketed piping and the operation of the cryogenic pumps and filtration system.

The Fermilab engineers have the most experience in designing and operating liquid argon systems in the US.  Pordes and Rebel are the primary contacts at Fermilab.  The LANL group is gaining experience in designing cryogenic systems and has been in close communication with Fermilab.   Mauger is the primary contact for the LANL efforts.

\subsection{MicroBooNE Design}
\label{sec:uBCryo}

The first cryogenic system for LArTPC detectors at the scale of 100~t built in the US is for the MicroBooNE experiment. It has not had any operational experience and  many of the fundamental designs are based on the success and lessons learned during the LAPD construction and operating experience. The system is currently being fabricated and will begin a partially outfitted test run Summer 2013. The major cryogenic subsystems of MicroBooNE are: $i)$ the initial cool down system, $ii)$ the purification system, and $iii)$ the nitrogen refrigeration system, as seen in  Fig.\ref{fig:uB}.  There is also a hot gas filter regeneration system. A brief description of the main components of each subsystem follows.

\begin{itemize}
\item Cool down System:  This system consists of a Fluitron diaphragm compressor, three pass aluminum plate fin heat exchanger, and a large capacity molecular sieve. The purpose of the system is to circulate warm gas to drive moisture off of surfaces in the vessel and then to cool the vessel and LArTPC to about 110~K over 2-3 weeks.  This slow cooling rate is designed to prevent thermal shock to the LArTPC wires.  {\bf Learned Lesson} The MicroBooNe cool down system was a major design task and cost for the experiment.  The cool down procedure of LArTPCs such as MicroBooNE's should be reevaluated for cost and efficiency.  Some possible alternatives to consider are: $i)$ designing the wire planes to withstand thermal shock so a slow cool down is not required, $ii)$ using heaters on the LArTPC frame to control temperature gradients, or $iii)$ using two phase nozzles during the initial purge that vaporize injected liquid, thus providing cooling without the thermal shock.
\begin{figure}[!htbp]
\centering
\includegraphics[width=3in]{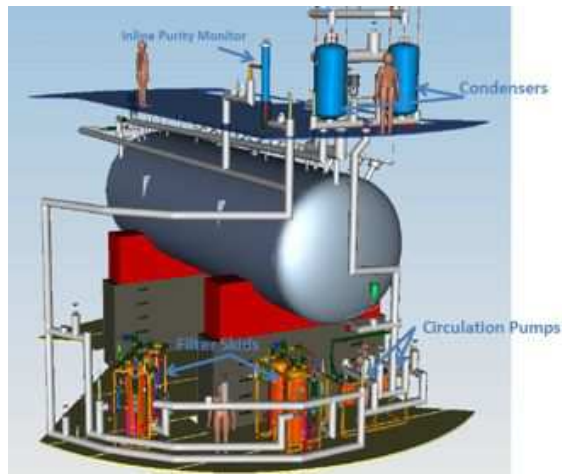}
\caption{{\scriptsize \sf  Rendering of MicroBooNE installation design. } }
\label{fig:uB} 
\end{figure}
\item Purification System:  This system includes the 35,000 gallon cryostat, two Barber Nichols circulation pumps that can run in parallel, two filter skids each containing a copper and molecular sieve, and a gas analyzer manifold for monitoring purity levels at different system locations. There are two pump skids and two filter skids to provide redundancy in the system.  The atmosphere in the cryostat is purged using the argon piston method demonstrated by LAPD.    The filter skids are designed from the LAPD experience.  They are heated with argon gas using a 3~kW circulation heater and regenerated with an argon and hydrogen gas mixture.  The argon is supplied by a 500 gallon dewar, and the mixture supplied by tube trailers. On site gas mixing systems for regeneration of the filters have proven to be far more cost effective over time than purchasing pre-mixed gas.
\item Refrigeration System: This system consists of two argon condensers each with a 9~kW capacity, phase separator for incoming liquid nitrogen and a 11,000 gallon nitrogen dewar.  The condensers are powered by liquid nitrogen from the storage dewar and implements a two coil system for heat exchange.  The system power can be regulated by operating one or both coils as needed.  As with the pump and filter skids, there are two condensers to provide redundancy in the system.
\end{itemize}

\subsection{LBNE Design}
LBNE is proposing to build a multi-kton detector in Lead, SD in a phased program.  It has developed a conceptual design which is highlighted here.  The LBNE cryostat will be purged following the method described in \S\ref{sec:LAPD}.  The proposed design uses the membrane cryostat technology commercialized for liquid natural gas (LNG) storage. This technology has proven reliable and is available from two companies in the world, GTT in France and IHI in Japan.  
\begin{figure}[t!]
\centering
\includegraphics[width=3in]{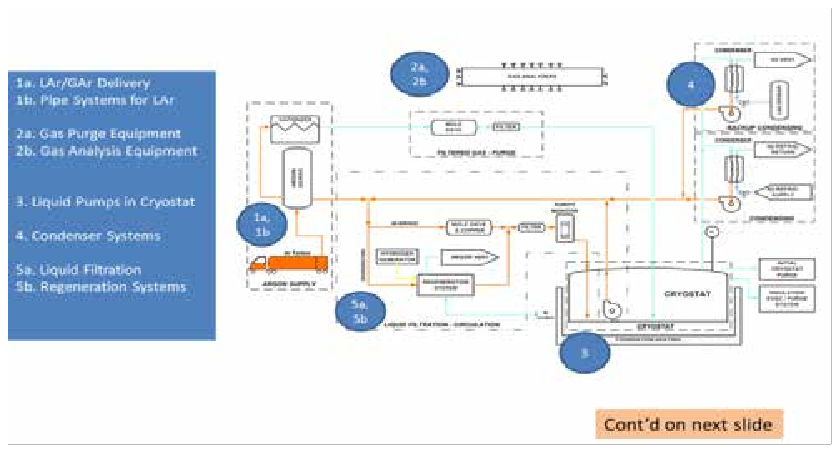}
\includegraphics[width=3in]{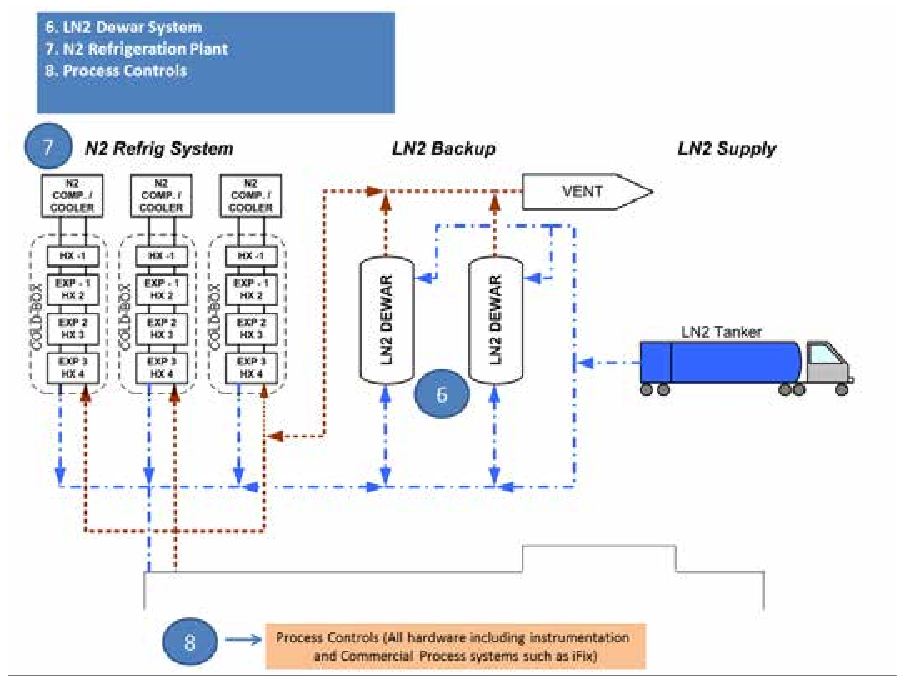}
\caption{{\scriptsize \sf [Top] Outline of Process Flow Diagram for 10-kton detector showing One of Two Cryostats. [Bottom] Process Flow Diagram for LN2 Supply } }
\label{fig:LBNE-1} 
\end{figure}

The design which is moving from concept to preliminary design stage can be broken into eight major sections as identified in Fig.~\ref{fig:LBNE-1}.  The initial phase requires 9.4 kton of liquid argon to fill each of two cryostats.  Cooling requirements suggest the need of three liquid nitrogen refrigerators at 55 kW each.  Under steady state conditions each cryostat`s boil-off argon will be liquefied by a single refrigerator and condenser pair and filtered before delivery back to the cryostat. During a period when one cryostat is full and purified to a level of  $<200$~ppt oxygen equivalent and the second cryostat is cooled down, the third refrigeration unit will be needed to provide extra cooling power.  

Gaseous and liquid argon filtering exists to maintain the high level of purity needed to achieve long electron long-time.  Mole sieve and copper are used in stainless vessels with an automated strategy for re-generation.  These are based on the LAPD design. Since LBNE is a 20 year experiment it was decided to use automated high purity, high temperature valves for heating of the beds to drive off oxygen and water.  These valves and the overall system will be managed by PLC logic to systematically control the process and limit the need for personnel to manipulate the process.  Argon gas is heated and mixed with hydrogen generated by a stand-alone unit.  The mixture is delivered to the filters, creating and releasing water which is exhausted from the vessels.  After driving off the impurities the same system is then cooled back down to operating conditions and placed back into operation.

Active liquid pumping of the argon in the cryostat will be continuous, requiring pump stations and towers to reside inside of the vessel.  This technology has been used in LNG membrane tanks, including the feed through ports for accommodating openings.

LBNE is building a membrane cryostat prototype.  This prototype will confirm the suitability of a membrane cryostat for LArTPCs.  It will also investigate the ability to make a vessel pure to $< 200$~ppt oxygen equivalent.  The prototype is anticipated that this vessel will be cooled down in summer 2013.

\subsection{LArIAT Design}
 LArIAT is an approved Fermilab test program experiment (T-1034) aiming to characterize LArTPC performance by placing a LArTPC in a charged particle test beam. This program is organized in two phases.  The first phase will place the ArgoNeuT detector in the MCenter beam line at the  Fermilab Test Beam Facility (FTBF),  while the second phase will build a larger LArTPC to be placed in the same beam line.  The LArIAT cryogenic facility will service both phases of the program. 

\begin{figure}[!htbp]
\centering
\includegraphics[width=3in]{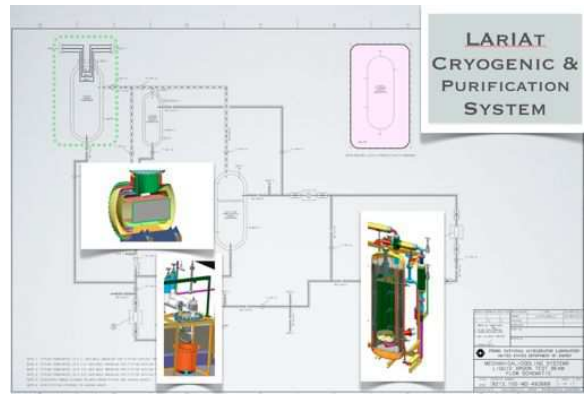}
\caption{{\scriptsize \sf  Schematic of the {\sf LArIAT} cooling and purification system. } }
\label{fig:LArIAT} 
\end{figure}

The LArIAT cryogenics system, shown schematically in Fig.~\ref{fig:LArIAT}, consists of two main elements: $i)$ the cryostat housing the active detector, and $ii)$ the new cryogenic system for argon cooling and purification forming a closed-loop system with the cryostat.   The cryostat for the first phase is a vacuum insulated dewar that contains $500$~L of ultra-pure liquid argon.  The original cryostat design relied on a gaseous argon purification system only that provided a full volume exchange every 7-8 days. The new cryogenic system has been designed to provide a full volume exchange every few hours. It is composed of a liquid nitrogen condenser for cooling and a forced recirculation network. Boil-off  gaseous argon is passed through a liquid nitrogen based condenser and the resulting liquid argon mixes with argon removed from the bulk volume.  This liquid argon is pushed through a  cryogenic pump housed in a vacuum insulated vessel to the downstream filtering system.  The filters are filled with a bed of molecular sieve and another bed of activated copper material to remove water and oxygen as in the LAPD system.  The purified liquid is returned to the main cryostat through the transfer line network. The circulation system is configured to ensure that the pump net positive suction will be adjustable for the different operations with both phases. The pump is known to have much larger capacity than required for the first phase and the high flow is accommodated by splitting part of it in a sub-loop circuit. The excess energy is removed through the high cooling power condenser unit.

\subsection{CAPTAIN Design}

Los Alamos National Laboratory is planning on building and running a LArTPC detector in a cryostat with a liquid argon volume of 7,500 liters or 10.5 tons. This system is known as CAPTAIN and its physics goals are described in \S~\ref{sec:CAPTAINTB}.  It is expected to be delivered to Los Alamos by the end of Summer, 2014.  The cryostat will be connected to a scaled down version of the MicroBooNE purification and recirculation system at Fermilab. The Los Alamos will use a single Barber-Nichols liquid argon recirculation pump and one purification tank that will contain both molecular sieve material to remove water and activated copper to remove oxygen.  Both will be regenerated at the same temperature using 2.5\% hydrogen in argon as done with LAPD. The major elements for this system are expected to be assembled Summer, 2014.

\subsection{Oxygen Deficiency Hazards}
\label{sec:ODH}
An oxygen deficiency hazard (ODH) occurs when gases such as argon or nitrogen displace oxygen in the air we breathe, potentially creating life threatening situations by lowering the oxygen concentration. LArTPC detectors require large quantities of liquid nitrogen and liquid argon, which if spilled or released indoors, can create a life-threatening ODH incident. For example if all the liquid argon in the $\sim38000$ gallon MicroBooNE cryostat were spilled and warmed up to room temperature there would be 4.2 million cubic feet of argon gas, vastly exceeding the size of the building housing the cryostat. To reduce the ODH risk, commercial oxygen monitors such as the MSA Ultima model are often used to measure the oxygen concentration in inside the buildings. For MicroBooNE, at oxygen concentrations below 19.5\% the emergency ventilation fan will pull 7500 CFM of a mixture of air and argon (nitrogen) from the lowest part of the building. The fan exhausts outdoors.  Alarms will also sound with sirens and strobes and the Fermilab Fire Department will be summoned. People inside the building should evacuate when there is an ODH alarm.  Before the cryostat can be filled with liquid argon an analysis must be made to determine the probability of a fatality due to ODH. If the probability is greater than one fatality every 10$^7$ hours, the average work place fatality rate, then there are special requirements for people entering the building.  These requirements are defined by the Fermilab safety manual, FESHM 5064~\cite{FESHM}. Formal ODH training and medical exam is required. Depending on the situation, there may be requirements for escape packs with breathing air, personnel oxygen monitors or multiple people in contact with each other.

{\bf Learned Lesson} Given the number of cryogenic systems planned or under construction, a standardized method for calculating ODH probabilities is needed.  Such standardization will facilitate the operational approve for these detectors and ensure safety throughout the program.

\section{TPC and High Voltage}
\label{sec:TPCHV}

\subsection{High Voltage Feedthroughs}
\label{sec:HV}

\begin{figure} [h]
\centering
\includegraphics[width=0.99\linewidth]{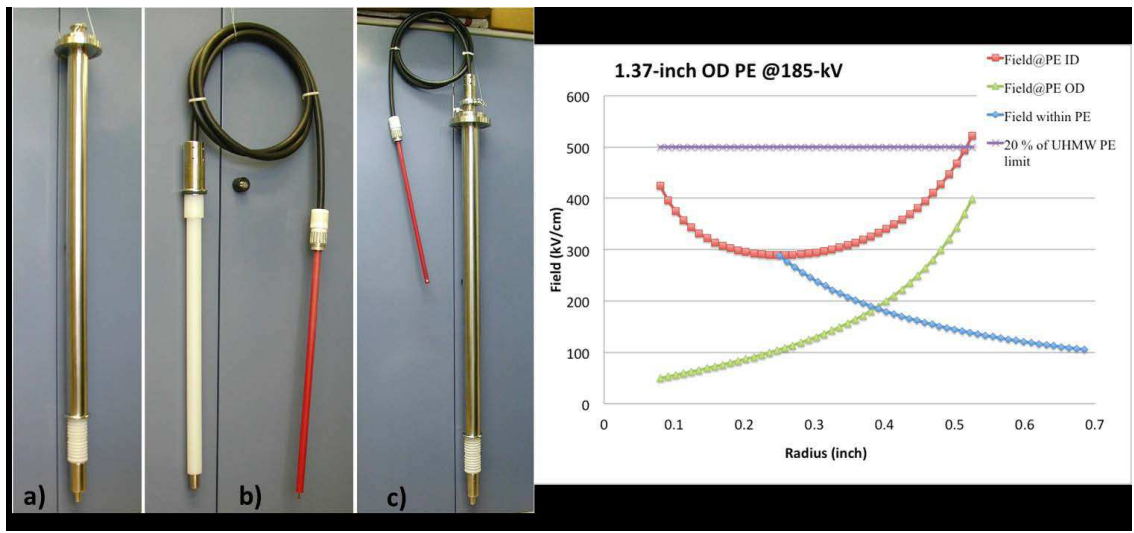}
\caption{Left) Componentes of the ICARUS feedthrough: a) HV feedthrough; b) cable plug; c) asssembly of HV feedthrough and cable plug. Right) Optimization of the electric field on the surface of the inner conductor (at 185 kV) for an outer conductor (at ground) of 1.37" diameter. The electric fields on the outer conductor surface and inside the UHMWPE are also shown.}\label{fig:HV}
\end{figure}

An essential element of all noble liquid TPCs is the generation of the high voltage (HV) to establish a stable and uniform drift field in the TPC sensitive volume. In case of the ICARUS LArTPC~\cite{Amerio:2004ze,Rubbia:2011ft}, with a maximum drift distance of $\sim1.5$ m, a cathode voltage of 75 kV is required for a nominal electric field of 500 V/cm. The UCLA group developed a cryogenic, custom made, HV feedthrough to carry the HV generated by an external power supply to the inside of the ICARUS cryostat on the cathode plane, as seen in Fig.~\ref{fig:HV}. The feedthrough, employing ultra-high-molecular-weight polyethylene (UHMWPE) as insulator, successfully operated up to a maximum of 150 kV~\cite{Amerio:2004ze}.  The groups in the US currently working on development of HV feedthroughs are UCLA (Wang) and Fermilab (Jostlein, Lockwitz).  

The LBNE feedthrough must provide a nominal voltage on the cathode of 185 kV. It is being developed by the UCLA group following the ICARUS design. Figure~\ref{fig:HV} also shows the optimization of the diameter of the HV inner conductor for an outer ground conductor of 1.37" diameter, in order to minimize the electric field strength in the insulator.

\begin{figure} [h]
\centering
\includegraphics[width=0.95\linewidth]{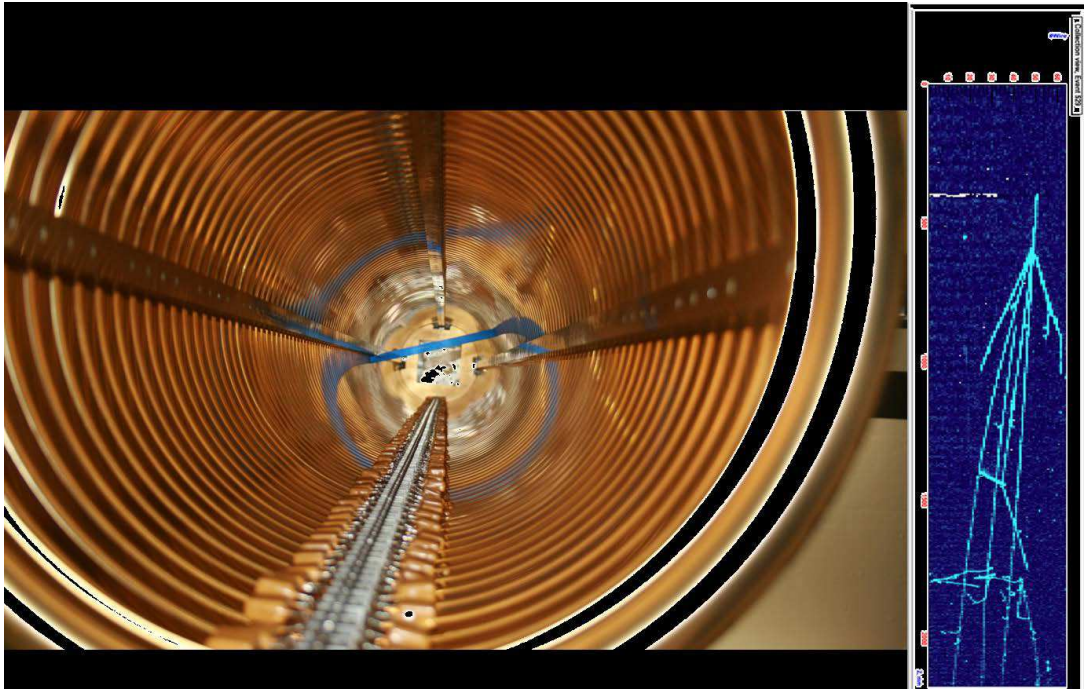}
\caption{Left) View of the ArgonTube TPC with a Cocroft-Walton voltage multiplier directly mounted on the field shapers. Right) Recorded cosmic ray interaction in ArgonTube.}\label{fig:ArgonTube}
\end{figure}

Longer drift distance LArTPCs are being considered, mainly in Europe, to minimize the number of electronic channels per unit volume. Full scale measurements of long drifts are necessary to assess the effects of signal attenuation, charge diffusion, accumulation of positive ions, and to test the HV systems.  The ARGONTUBE LArTPC~\cite{Ereditato:2013xaa} has a vertical drift of 5~m and is shown in Fig.~\ref{fig:ArgonTube}.  It is presently being operated at the University of Bern. Cosmic muons tracks have been detected with drift distances of up to 5 m. The ARGONTUBE field cage is assembled as a set of 125 field shaping rings mounted in a column with 4 cm pitch and 5 mm clearance between rings. 

Cockcroft-Walton (or Greinacher) voltage multipliers immersed in liquid argon, with the different stages of the multiplier directly connected to the field shaper electrodes, have been proposed by the ETH Zurich group~\cite{Horikawa:2010bv, Badertscher:2012dq} as an effective way to reach very high voltages directly in liquid argon, eliminating the need for a HV feedthrough.  ARGONTUBE utilizes a voltage multiplier with 125 stages, driven by a maximum ac voltage of Vpp=4 kV peak-to-peak, for maximum HV of 500 kV on the cathode. The voltage multiplier circuit, placed inside the field cage in order to minimize the electric field at the surface of its small structures, slightly distorts the drift field. The device has been stably operated up to a voltage of 120~kV. Discharges encountered at higher voltages are under investigations. A system composed of two 4 cm diameter spherical electrodes, with a positioning accuracy of $5 \mu$m, is presently being employed at the University of Bern to study the breakdown voltage in liquid argon.

The experience of building stable HV feedthroughs is limited to a few individuals.  The longest drift distance LArTPC in the US, LongBo, currently operates at 70~kV but has a requirement of 100~kV for its 2~m drift distance.  The individuals at the workshop who have experience building HV feedthroughs agreed that making stable feedthroughs is very difficult and there is no set recipe for doing so. The field would benefit from a more extensive R\&D to develop stable feedthroughs, exploiting synergies with noble liquid dark matter detectors.   

\subsection{TPC Construction}
\label{sec:TPC}

\begin{figure} [h]
\centering
\includegraphics[width=0.90\linewidth]{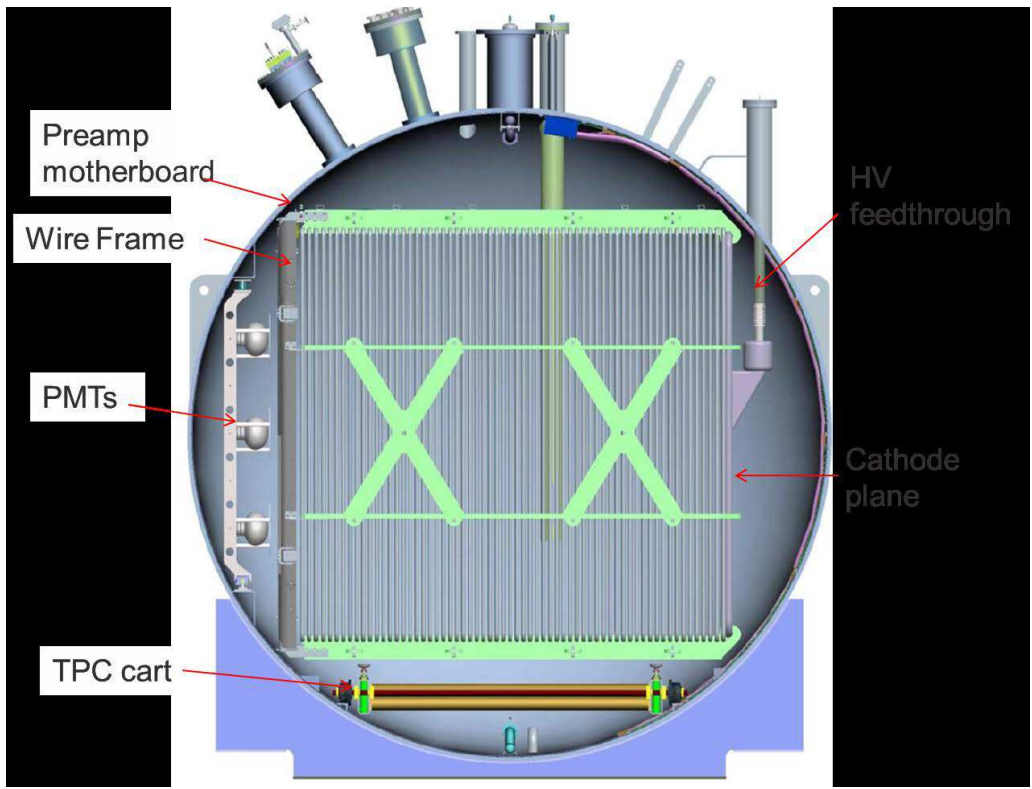}
\caption{Overview of the MicrBooNE TPC.}\label{fig:MicroBooNE_TPC}
\end{figure}

The construction of TPCs is being done in the context of the experimental program.  MicroBooNE, with its 2.5~m $\times$ 2.5~m $\times$ 10~m TPC, shown in Fig.~\ref{fig:MicroBooNE_TPC} is the first 100~t scale LArTPC to be built in the US.  It represents several innovations in LArTPC design, including
\begin{itemize}
\item a maximum drift length of 2.5 m, a substantial increase relative to the 1.5m drift of ICARUS;
\item vertically oriented collection wires providing good dE/dx sampling for tracks along beam direction;
\item wires are mounted on rigid, adjustable frames, with no spring loaded wire tensioning structure; and
\item copper plated wires to improve conductivity and reduce noise.
\end{itemize}

The construction of the MicroBooNE LArTPC and has provided several practical {\bf Learned Lessons} during its construction, some simple but not obvious, others more complex. 

One feature of the anode frame design teaches a lesson that should not be forgotten for future designs of large TPCs: the frame itself, which holds the rigid wire tensioning structure, is a C-channel created by bending a flat piece of 5/8-inch thick stainless steel plate stock. Achieving the required tolerances on a tight bend radius in a large piece of metal is difficult for most machine shops.  The consequence of not meeting the tolerances in the case of MicroBooNE was that the anode frame pieces did not fit together to form a perfect rectangle, and the pieces had to be coerced into the proper shape during assembly by means of a force multiplying come-along tool. For frame design, the simpler the better, in terms of both machining and of assembly.

The machining and cleaning of the different parts of the TPC are important aspects not be neglected. The MicroBooNE field cage is built out of 1" diameter stainless steel tubes, with holes drilled every 6 inches along the length of the pipes to improve argon flow. There was no indication to the machine shop that each hole should be carefully deburred, so the holes had to be deburred by hand upon arrival at Fermilab.  That process was both tedious and time-consuming step. Each part of the TPC must be cleaned before assembly. Any hollow tubing or pipe that will have its ends permanently capped, but has other holes along its length should be cleaned thoroughly by the machine shop {\it{before}} the ends are capped, otherwise it is very difficult to remove metal shavings and dirt from the hollow piece.

\begin{figure} [h]
\centering
\includegraphics[width=0.9\linewidth]{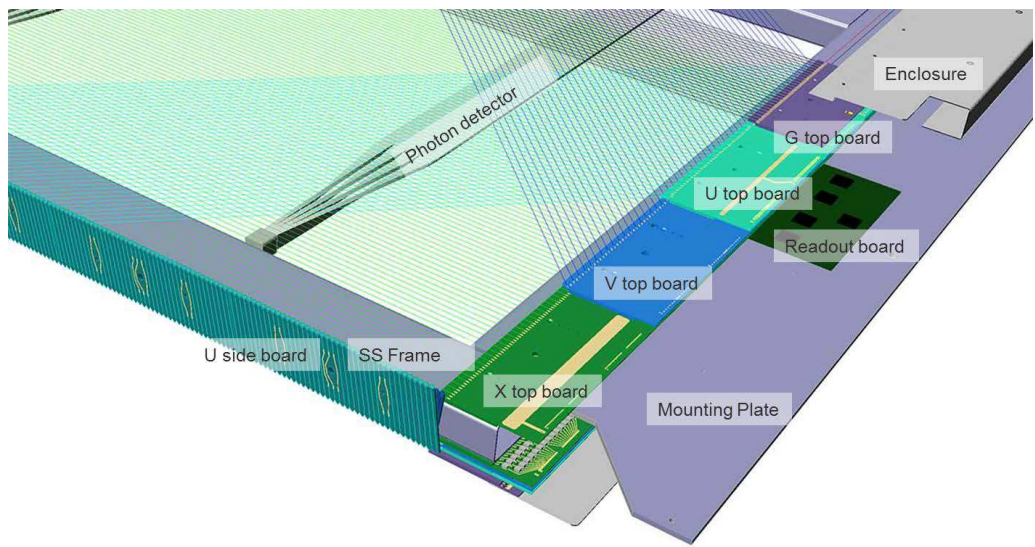}
\caption{Schematics of the LBNE Anode Plane Assembly (APA).}\label{fig:LBNE_TPC}
\end{figure}

The LBNE TPC will be built in a new manner that has not been experimentally used.  The detector is built as a three-dimensional array of double TPC cells with 2.3 m drift.  The TPC cells are constituted by a double sided anode wire plane assembly (APA) in the middle and a cathode plane assembly (CPA) on each side. Figure~\ref{fig:LBNE_TPC} shows the details of an APA module. 
The LBNE design, in order to reduce the complexity and cost of a multi-kiloton detector, introduces new concepts with respect to the MicroBooNE TPC:
\begin{itemize}
\item wire planes have a double sided wire wrapping configuration;
\item the use of CuBe wires allows soldering for electrical connections; the wires are mechanically held by epoxy;
\item a grid, placed in front of the wire planes, enhances the signal of the first induction wire plane and protects the electronics from electrostatic discharges;
\item the field cage is constructed from printed circuit boards.
\end{itemize} 

Both APAs and CPAs are 2.5 m wide (in the beam direction) and 7 m high. On each APA, four planes of wires cover each side of a frame. The wires on the outermost plane are oriented vertically and are not connected to the readout electronics. They shield the three inner sense wire planes, which have a nominal wire pitch of 4.5 mm. The inner planes are labeled, along the direction of the electrons drift, as: U\&V induction planes, and Z collection plane. The U\&V wires are positioned at $\sim \pm45^\circ$ to the vertical, and the Z wires are vertical. The wires on the U\&V induction planes are wrapped in a helical pattern around the long edges of the wire frame, so that the readout electronics can be located on the top or bottom of the TPC. This arrangement results in 7-m-long collection-plane wires and 10-m-long induction plane wires. To avoid ambiguities in track reconstruction, due to the wrapping of the U\&V wires, the wire angles of U\&V planes are set slightly differently: $45^{\circ} \pm \delta$, such that the same three wires only cross once.

The operation of a LArTPC in double phase (liquid-vapor) opens the possibility of amplification of the ionization charge in the pure argon vapor phase.  Ionization electrons are drifted towards the liquid surface at the top of a vertically standing cryostat, then extracted in the pure argon gas phase and there amplified by thick macroscopic GEMs, known as Large Electron Multipliers (LEMs), or THGEMs. These devices, manufactured with standard PCB techniques, are thought to represent a robust and economic way to realize large area detectors, suitable for cryogenic operation.

An active R\&D program is being conducted by the ETH Zurich group. The main motivation is to improve the quality of the ionization charge signals of a LArTPC; flawless reconstruction, including reconstruction of low energy supernova neutrinos, requires excellent signal to noise ratio and the collection view provides the best imaging. This technique would somewhat compensate for the diffusion and charge attenuation over several meter drift distances and it could offer interesting applications for detection of below minimum-ionizing-particle signals, e.g. dark matter.

\begin{figure} [h]
\centering
\includegraphics[width=0.95\linewidth]{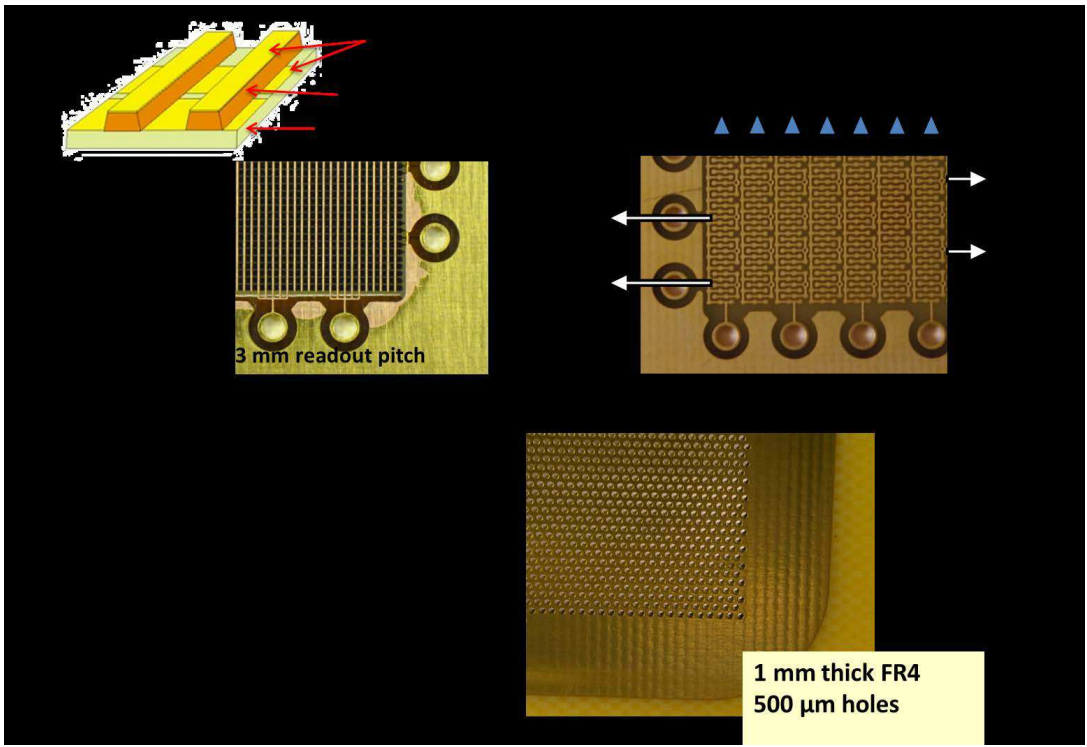}
\caption{Large Electron Multiplier (bottom) and different schemes for the two-dimensional projective anode readout.}\label{fig:LEM}
\end{figure}

\begin{figure} [h]
\centering
\includegraphics[width=0.9\linewidth]{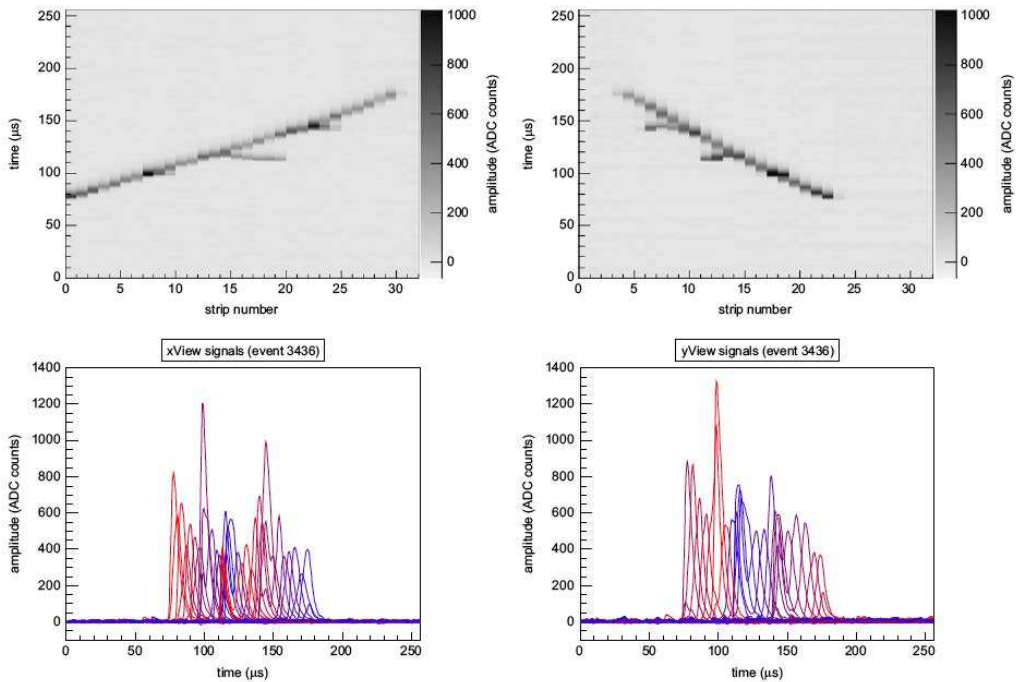}
\caption{Othogonal views of a cosmic muon track and associated signals for a LEM-TPC.}\label{fig:LEM_TPC}
\end{figure}

A LEM TPC prototype was successfully tested~\cite{Badertscher:2010zg} with a $10\times 10$ cm$^{2}$ LEM, operated with an electric field of ~35 kV/cm. The ionization charge, amplified by the LEM, is drifted to a two-dimensional projective readout anode, which provides two orthogonal views with 3 mm spatial resolution as seen in  Fig.~\ref{fig:LEM}. An effective amplification of $\sim 30$ of the released ionization charge has been achieved without any degradation in resolution, providing a S/N ratio larger than 200 for minimum ionizing particles, as seen in Fig.~\ref{fig:LEM_TPC}.

Recently a LEM LArTPC with a readout unit of $76 \times 40$ cm$^2$ has been successfully tested~\cite{Badertscher:2013wm} with an effective gain of $\sim 14$. Work is in progress to optimize the extraction of the ionization electrons from the liquid and to simplify the two-dimensional anode readout. The stability of the achieved gain over long periods needs to be assessed and it is presently being investigated.

\section{Electronics, Data Acquisition, and Triggering}
\label{sec:Electronics}
\subsection{Front End Electronics}
\label{sec:FEE}
Several groups are contributing to the development of front end electronics for use in LArTPCs.  Those groups include Brookhaven National Laboratory (Chen, De Geronimo, Nambiar, Fried, Ma), Fermi National Accelerator Laboratory (Deptuch, Hoff), Michigan State Univiersity (Bromberg, Edmunds, Shooltz) and Southern Methodist University (Gui, Ye, Liu).  These efforts are closely coordinated under the LBNE Project, and are working toward a common system based on the BNL ASICs and MicroBooNE construction experience.

\begin{figure}
\centering
\includegraphics[width=3.5in]{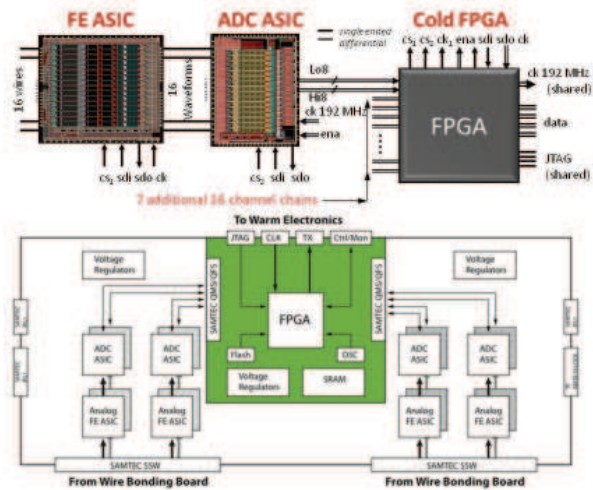}
\caption{\label{fig:35tonelectronics} Cold electronics system for the LBNE prototype detector.}
\end{figure}

The BNL group is developing a complete signal processing chain that operates in liquid argon~\cite{Radeka,Thorn}.  They have developed device models and design rules for a 180nm TSMC CMOS analog process that can operate at all temperatures from above room temperature to below 77K.  A low noise, low power 16 channel front-end analog ASIC is now available~\cite{Geronimo}.  The functional parameters of this circuit are fully programmable to select gain, shaping time, pedestal level, and output coupling.  Gain matching and calibration capacitance is matched to better than 1\% channel to channel across 90\% of all chips.  Power consumption is ~6 mW per channel.  At present, 8256 channels of cold front end electronics using this ASIC are being installed in MicroBooNE~\cite{Chen} and it will also be used for the CAPTAIN TPC being constructed by LANL, and the LArIAT TPC being constructed at Fermilab. A low power, clock-less, 16 channel, 12 bit ADC is also in development in this same CMOS process for the LBNE far detector.  The functional lifetime of these two ASICs has been demonstrated to be in excess of 20 years. The operation of commercial FPGAs at 77K is being investigated for use in processing of the digitized signals for buffering, zero suppression, data compression, multiplexing and transmission out of the cryostat in large TPCs such as the LBNE Far Detector.  Figure~\ref{fig:35tonelectronics} indicates the features of such a system being constructed for the prototype being constructed for the LBNE Project.  Several devices and conditions for operation have been identified for use at 77K.  Studies have begun to understand operational lifetime at 77K.  Studies of long term operation of commercial voltage regulators at 77K have identified several suitable devices and indicate that the functional lifetimes can be in excess of several years.  

The SMU group is looking at the operation of commercial FPGAs at 77K, in parallel with the efforts of the BNL group.  They have identified devices and conditions for operation at 77K, and made measurements hinting that operational lifetimes can be in excess of several years at 77K.  In addition to this work, they have studied the operation of optical transmitters at 77K.

The Fermilab group is pursing a related path by developing device models and design rules for a 130nm Global Foundries CMOS digital process that can operate at all temperatures from above room temperature to below 77K. Measurements of device lifetimes as a function of drain-source voltage indicate that digital CMOS ASICs can have operational lifetimes of decades in liquid argon. This technology would be available for implementing the data sparsification, packing, and multiplexing that are presently envisioned to be performed by commercial FPGAs, if it is found that such FPGAs cannot be shown to have functional lifetimes of decades in liquid argon.  The SMU group is contributing to this work by characterizing the 130nm Global Foundries CMOS digital process for operation in liquid argon.  They are measuring device parameters and measuring lifetime as a function of voltage stress.

The Michicagn State group designed 480 channels of warm discrete JFET electronics for the ArgoNeuT LArTPC, which is the only one to take data in a neutrino beam in the US. Subsequently they constructed and successfully operated 144 channels of cold hybrid CMOS front end electronics for a 30~cm drift distance TPC at Fermilab.  They have also implemented 16 channels of readout using the BNL front end ASIC on a 200~cm drift distance TPC, which is now being operated in LAPD.  

\subsubsection{Design Considerations}

All the efforts mentioned above must account for the fact that the lifetime of in-cryostat electronic systems for use in very large LArTPCs must be greater than the expected life of the experimental program.  The expense of replacing or storing the liquid argon is so great that the threshold for replacing or repairing cold electronics is very high.  Therefore, all electronic components for use in liquid argon should have lifetimes known to be greater than 20 years.  The effect that determines lifetime at cryogenic temperatures is hot carrier injection into the gate interface.  For custom ASICs long lifetimes can be ensured by proper design rules, and verified by tests on individual, isolated devices.  For commercial devices, especially FPGAs, careful selection and testing may be able to demonstrate long cryogenic functional lifetimes.  This understanding must be developed.  

The formation of bubbles in liquid argon increases the probability of breakdown in high field regions.  Conditions for boiling in liquid argon coming from the power density and surface conditions, for example, need to be determined.  The devices and techniques used need to prevent or control bubble formation in liquid argon.

The final consideration for development of front end electronics is the impact of external noise on the systems.  External noise sources from liquid argon flow and acoustic coupling as well electromagnetic noise from the liquid argon purification system must be minimized.  The bandwidth to retain high spatial and timing resolution must be optimized.

\subsubsection{Synergies with Dark Matter Experiments}

Dark matter experiments may benefit from the development of cold front end electronics for both the TPC and light collection systems.  Specifically, the development of low noise and low power consumption cold electronics could be of benefit, although these experiments inherently have fewer channels and higher rates per channel.

\subsection{Data Acquisition}

The needs of a data acquisition (DAQ) system are highly dependent on the experimental goals.  As such, DAQ development in the US has largely been in support of the MicroBooNE and LBNE experiments.  

The MicroBooNE work is being done by Columbia/Nevia Laboratory (Camileri, Karagiorgi, Sippach, Chi), Yale (Church), BNL (Chen) and Fermilab (Baller).  This collaboration has designed and constructed the DAQ system to collect data from 8256 wires sampled at 2MHz and 30 PMTs sampled at 64 MHz.  The system is implemented on a custom ÒTPC/PMT Readout BoardÓ with ADCs, FPGAs and buffer memory for event building and trigger generation. Each readout crate contains one transmit module (XMIT) to collect data from the TPC or PMT readout boards through the backplane and to transmit data to a DAQ PC in each crate through fiber optic links and a PCI Express card. The DAQ utilizes ten readout crates, and therefore ten transmit modules, ten PCs, and 31 PCI Express cards for the entire detector.  The MicroBooNE detector operates in two primary modes, one each to collect accelerator beam and supernova data, each with its own data stream. The beam triggered data stream rate is 1.2 MB/s per crate with events built and stored from crate data. The supernova stream rate is 50 MB/s per crate to the DAQ, with a much smaller rate to permanent storage.  These data are queued until expiration or receipt of a supernova ÒtriggerÓ from SNEWs. 

The Fermilab (Biery, Bowden, Kwarciany, Rechenmacher, Johnson), Indidna (Urheim), and SLAC (Convery, Graham, Herbst, Huffer) groups are proposing three competing solutions for the LBNE Far Detector DAQ system.  The original proposal is to use the NO$\nu$A Data Concentrator Modules (DCMs) with the addition of an FPGA based data interposer board between the front end and the DCMs.  This is a well developed solution, but perhaps too slow and somewhat inflexible in matching rates per channel and number of channels to the LBNE system.  A more modern and flexible system in the early stages of development has been proposed by Fermilab Computing Division, using commercial PCIe cards, which are available with up to 138 differential LVDS signals routed to FMC connectors. It would use the Fermilab designed ARTDAQ software toolkit for data transfer to disk. The third option under consideration for the LBNE TPC is to adopt the SLAC system based on ÒReconfigurable Cluster ElementÓ (RCE) packaged in an ATCA crate.  This system is being used for several projects, such as LSST. Both the RCE and the commercial PCIe system are designed for high speed connections to intelligent data sources, such as the FPGAs planned for the LBNE front electronics system.  The SLAC RCE system has been chosen for the 35 ton test LArTPC, to be operated in late 2014, with the DCM solution as a backup, in case the RCE solution cannot be adapted in the short time available.

The efforts of the various LBNE groups involved in DAQ development are essentially competitive, with each system having been, or being, developed for uses other than in LArTPCs; NOvA for the DCMs, LSST and LCLS for the RCEs, and Mu2e for the commercial PCIe based DAQ.  This competition is understandable at this stage in the development of the LBNE project as it does not yet have a baseline design for the DAQ.

\subsubsection{Design Considerations}

One key consideration for all DAQ systems is that the systems must incorporate algorithms for zero-suppression and data compression that preserve the physics content of the TPC data.  The need for these algorithms arises from the large data rates and volumes that result from untriggered operation of very large active volume LArTPCs with high intrinsic spatial resolution. The data rates indicated above for the MicroBooNE system illustrate the need for such algorithms.  This effort will also involve matching the number of channels and rates per channel between the front end and DAQ.

Studies are also needed of the interaction of the electronic performance on the data analysis software being developed for LArTPCs.  Models of signal formation, noise, and electronic response need to be refined in the simulations and analysis.

\section{Scintillation Light Detection}
\label{sec:Photons}
The efforts at understanding photon detection in LArTPCs are aimed at supporting the current generation of experiments in the planning or construction phase, MicroBooNE, LBNE, and LArIAT.  These efforts are led by MIT (Conrad, Jones, Katori) for MicroBooNE;  Indiana University (Mufson), Colorado State University (Buchanan) and Lawrence Berkely National Laboratory (Gehman) for LBNE; and Yale (Cavanna, Szelc), Fermilab (Kryczy\'nski) and the University of Chicago (W. Foreman, D. Schmitz) for LArIAT.

There are many synergies within these groups since these efforts are all aimed at developing the technologies needed to detect scintillation photons in liquid argon for neutrino physics.  

\subsection{Effects of Nitrogen Impurities on Attenuation Length at 128 nm}
\label{sec:scintN2}
Water and oxygen are well-understood as important impurities for electron drift in LArTPCs.  These impurities can also affect also the collection of the scintillation light at sufficiently large concentrations. Oxygen at concentrations above 100~ppt was found to cause both quenching of the long-lived Ar$^*_2$ excimers before scintillation light emission and absorption of the emitted VUV photons~\cite{warpoxygen}.  Long drift distance detectors require smaller concentrations of oxygen in order to collect ionization electrons, meaning that for these systems the effect of oxygen on the scintillation light production is controlled by the requirements for drifting the ionization electronics.

Nitrogen contamination is usually assumed to be less harmful in liquid argon since it does not affect the electron drift.  For scintillation light collection N$_2$ quenching becomes relevant at concentrations above 0.5 ppm~\cite{warpnitrogen}.  The N$_2$ photo-absorption cross section at 128 nm is small, however, it was noted that for extended detector dimensions light reduction due to absorption over long distances may not be negligible. The absorption length measurements as a function of nitrogen concentration described here are the first to be made at 128 nm.

The absorption length studies were carried out with the MicroBooNE vertical slice setup, which mocks up one slice of the MicroBooNE optical system in a test dewar at Fermilab filled with liquid argon that was filtered using the MTS argon delivery system described in \S~\ref{sec:MTSP}.  A $^{210}$Po alpha particle source was used to excite the scintillation photons.  Parts per million amounts of nitrogen were injected into the liquid from a gas canister, charged to a known pressure.   By making two measurements at different distances from the PMT, the effects of nitrogen on attenuation length can be separated from those due to quenching.  

\begin{figure}
\centering
\includegraphics[width=3.5in]{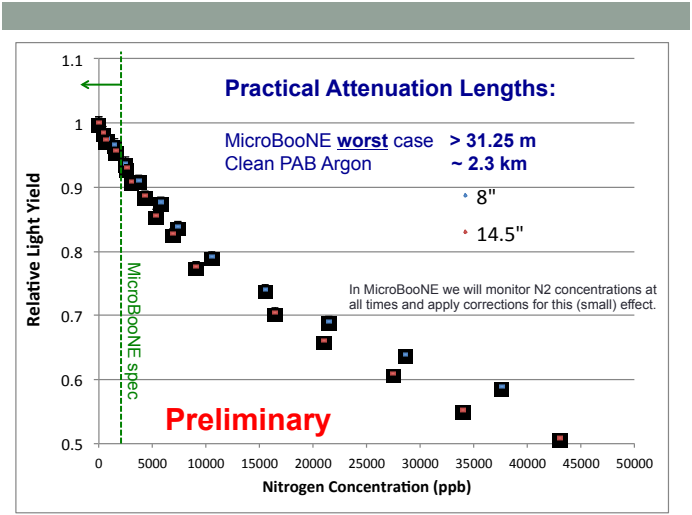}
\caption{\label{fig:n2impurities} Attenuation of 128 nm light as a function of nitrogen concentration in parts per billion.}
\end{figure}

The results shown in Fig.~\ref{fig:n2impurities} indicate that the attenuation length of ``clean'' liquid argon is $\sim$2.3~km and only drops to $\sim$30~m for a concentration as large as 30-40 ppm.  Monitoring the nitrogen content will minimize this small effect in MicroBooNE.  {\bf Learned Lesson} In future large LArTPC experiments, it will be important to keep the nitrogen content in the liquid argon to $<1$ ppm.  Keeping the nitrogen impurities below this concentration will assure that the reduced liquid argon attenuation length will not become an issue.

\subsection{Comparison of bis-MSB and TPB as VUV Waveshifters} 

The photon detection systems for liquid argon (128~nm) and liquid xenon (175~nm) detectors typically employ waveshifters to down convert the VUV photons into the optical for significantly more cost-efficient read out with optical photodetectors.  Detection of VUV photons directly is a more challenging problem.  For large LArTPC detectors, where photocathode coverage is prohibitively expensive, the proposed solution is to coat waveguides or fibers with waveshifter and then route the converted VUV photons to optical photodectors at the ends.  In more modest LArTPCs, where reasonable photocathode coverage can be achieved, the photodetectors view the scintillation photons behind wavelength shifting plates or waveshifter deposited directly onto the photodetector.  In dark matter experiments, the the waveshifter is often applied directly to the photodetector. 

\begin{figure}
\centering
\includegraphics[width=3.5in]{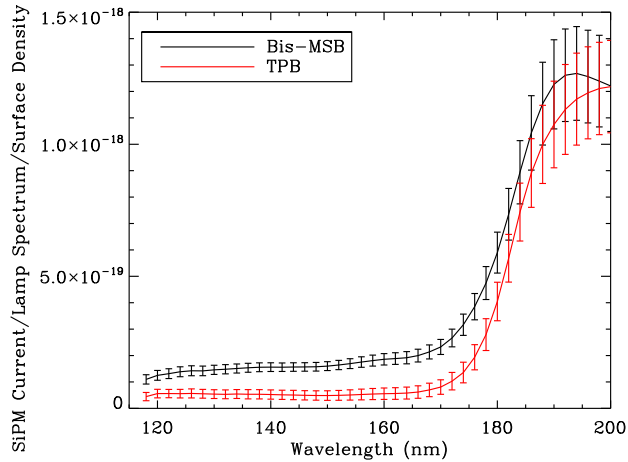}
\caption{\label{fig:tpbbismsb} Attenuation of 128 nm light as a function of nitrogen concentration in parts per billion.}
\end{figure}

TPB is the most widely used waveshifter to convert VUV scintillation photons from liquid noble elements.  At IU, CSU, and LBNL, investigations are underway to determine whether bis-MSB can provide a cost-effective alternative to TPB, especially important for large detectors.  The measurements made at IU in a VUV monochromator with cast acrylic light guides, and shown in Fig.~\ref{fig:tpbbismsb}, demonstrate that bis-MSB and TPB are equally effective as waveshifters.  Measurements at LBNL show that bis-MSB and TPB are equally efficient when deposited on waveshifting plates.  CSU is beginning studies to compare the waveshifting properties of bis-MSB and TPB deposited on and within polystyrene fibers.  

\subsection{Cast Acrylic Light Guides and Fibers} 

LBNE is investigating two technologies for its photon detector system for LArTPCs.  At IU the technology being investigated uses cast acrylic light guides with waveshifter embedded on its surface as the detector element to be assembled into a larger system.  This design was first described by the MIT MicroBooNE photon detector group~\cite{mitlightguide}.  Cast acrylic was found to have significantly better light transmitting properties than cast polystyrene, extruded polystyrene, and extruded acrylic when measured at 420 nm.  Tests at IU have shown that Lucite UTRAN cast acrylic bars have the longest attenuation length ($> 2$~m) when compared with other commercially available UVT plastics.  UVT plastic is superior to all UVA plastics tested.  The waveshifter is deposited onto the plastic's surface by an automatic, computer-controlled spraying apparatus.  The waveshifter is then embedded into the plastic by rapidly heating to high temperature and then cooling in a water bath.  The heating and cooling operation must be fast or the bars warp, thereby significantly degrading their properties as light guides.

The CSU group is actively engaged in testing polystyrene fibers embedded with waveshifter in their cores as a photon detectors for LArTPCs.  In this technology, the fibers directly capture the VUV scintillation photons and channel the waveshifted light to the photodetectors.  There are many advantages to fibers, including their well-understood production and their cost-effectiveness.  Fibers also have a long history successful performance in HEP experiments.  The challenge here is to embed the waveshifter and use them in a cryogenic environment.

The LBNL group has been investigating the commercial production of cast polystyrene light guides with waveshifter.  In these prototypes, the waveshifter, TPB or bis-MSB, is added to the polystyrene volume before the light guides are cast.  Currently samples of cast light guides with both waveshifters that have been produced by Eljen Technologies are under investigation at the VUV monochromator setup at LBNL.

\subsection{MPPCs/SiPMs as Photosensors }

Indiana University has been testing Hamamatsu MPPCs and SensL SiPMs on their prototye cast acrylic light guides.  Single photoelectron peaks have been seen for both devices in tests in liquid argon temperatures when used in conjunction with the MicroBooNE amplifier and shaper circuit.  As expected the noise decreases significantly at liquid argon temperatures compared with room temperature.   When the light guides are immersed in liquid argon and illuminated by an $^{241}$Am source, the expected Poisson-distributed signals are seen.  When a cosmic ray hodoscope provides the trigger, several very strong signals from muons traversing the liquid argon have been observed.  For large LArTPCs, there is significant promise for a photon detector system designed around cast acrylic light guides coated by a VUV waveshifter and read out by MPPCs or SiPMs at their ends.

\subsection{Testing Cryogenic PMTs in an Open Dewar}

\begin{figure}
\centering
\includegraphics[width=3.5in]{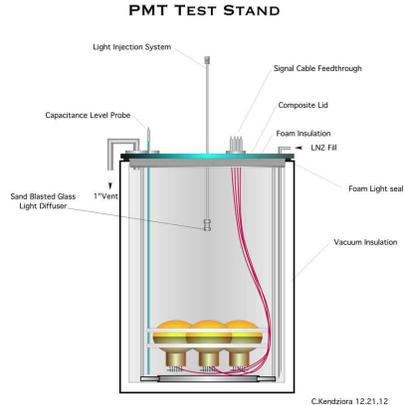}
\caption{\label{fig:pmtteststand} Schematic of test stand for characterization of cryogenic PMTs.}
\end{figure}

The MIT group (Kartori) designed the test stand in Fig.~\ref{fig:pmtteststand} to characterize the MicroBooNE cryogenic 8-inch PMTs.  It is based on an open dewar concept and is not a cryostat.  The test stand is a demonstration of a modest cost cryogenic system that can easily be duplicated for moderate cost and will be valuable even for small university groups.  This system can effectively test key features of cryogenic PMTs, including, dark current measurement, and long term stability.  Because it is an open dewar system, however, impurities are not well controlled.  Consequently, the system cannot test subtle effects like the response to radioactive backgrounds.  Nevertheless, this system is still quite useful in testing key properties of cryogenic PMTs in a cost-effective way.

\subsection{Test Facilities}

The Colorado State University group is currently implementing a liquid argon cryogenic detector test facility.  It is built around a 500~L dewar and is designed to accommodate 2.2~m long photon detector components.  It is equipped with a 1~t crane to allow components to be quickly and easily added or removed.  In case of a spill, there is a large catch basin and a venting system to push gaseous argon outside.  This facility is now installed and undergoing commissioning tests.

Fermilab also has a test facility for photon detectors in liquid argon.  It has a dewar with a 20~inch inner diameter and 60~inch depth of argon into which various photon detection schemes can be placed.  The dewar is connected to a source of purified argon, a condenser for the boil off argon, and several gas analyzers to determine concentrations of contaminants such as oxygen, water and nitrogen.  This facility is called Tall-Bo.

\subsection{Calorimetric Reconstruction}
\label{sec:LArIATPhot}

The LArIAT test beam experiment described in \S~\ref{sec:LArIATTB} will test the importance of scintillator information in event reconstruction in LArTPCs.  The value of calorimetric reconstruction using scintillator information has already been demonstrated in dark matter experiments~\cite{lightcal}.  The LArIAT experiment consists of  2 PMTs, operating at cryogenic temperatures, and a highly reflecting foil, coated with a thin TPB film, on the inner surfaces of the field cage of the LArTPC.  The VUV scintillation photons from events in the cryostat are shifted into PMT observable photons after hitting the TPB and are then reflected from the mirror surfaces beneath.  LArIAT will have a Òdark matter-likeÓ light readout system for calorimetric reconstruction and pulse shape discrimination.  Tests with a prototype detector at the University of Chicago are scheduled to begin in Spring, 2013.  If the calorimeteric energy reconstruction with LArIAT is successful, it may develop into the standard design for future LArTPCs.

\subsection{Outstanding Issues to be Resolved}

Perhaps the most outstanding issue to be resolved has to do with environmental effects on VUV waveshifters.  In particular, recent work at MIT has shown that not only does UV radiation degrade the efficiency of TPB, but this degradation is irreversable.  The IU group have found that UV light similarly degrades bis-MSB.  The evidence complied at MIT suggests that degradation may in fact affect most photosensitive waveshifters.  The importance of this issue stems from the fact that large LArTPCs will require large numbers of photon detectors and detector assembly techniques will therefore need to be developed so that manufacture and installation of these photon detectors minimally degrades the waveshifters by commissioning.  Further, it is clear that the photon detectors need to be shielded from sunlight and room light at all times.  The IU group is testing assembly techniques that block the light guides from UV radiation from room lights.

A second important issue is the need to quantify the allowable concentration of nitrogen in liquid argon for successful operation of photon detectors in LArTPCs.  As shown by the MIT group (Jones), nitrogen not only results in a shorter liquid argon attenuation length for scintillation photons, it also quenches the production of scintillation light altogether.  While allowable concentrations of water and oxygen have been determined, no precise requirements presently exist for nitrogen.  It will be very important to establish these requirements going forward.

A third issue to be resolved concerns production techniques for the liquid argon photon detectors.  For large LArTPC detectors like LBNE, a large number of photon detectors will be needed.  How is the waveshifter to be incorporated into the plastic, whether in light guide bars or fibers?  Currently several techniques are being investigated.  For bars, the waveshifter can be embedded into the plastic by mechanically spraying it on the surface and then melting it into the outer surface of the bars (IU).  The waveshifter can be also be painted onto the surface in a polystyrene plus solvent coating (MIT) that embeds the waveshifter into the plastic.  Finally, the waveshifter can be mixed into the plastic before it is cast into bars (LBNL).  With that technique, waveshifter is distributed throughout the volume, although the penetration depth of the 128 nm liquid argon scintillation photons into the plastic is very short.  For embedding the waveshifter into fibers (CSU), two techniques are actively being investigated.  In one technique, the waveshifter is embedded into the cladding and waveshifted light is trapped and channeled by the polystyrene core.  In the second, the waveshifter is mixed into the polystyrene core.  In this technique, the VUV photons are transmitted through a thinned cladding and then waveshifted in the core.  

\subsection{Future Directions}

There are two vital directions for the future.  First it is essential to learn how to integrate the photon detectors to the LArTPC detectors.  At present MicroBooNE is integrating their photon detection system into the detector construction now in progress.  The MicroBooNE system consists of wavelength shifting plates in front of 8'' PMTs.  The system design is well-developed and the components are currently undergoing quality control testing.   The MicroBooNE experience will clearly benefit future LArTPCs in the design phase.  For LBNE the integration will be tested in the 35 ton prototype test currently scheduled for Fall, 2014.  In these tests a small scale prototype of the full LBNE LArTPC will be integrated and tested.  All the issues to be resolved discussed in the previous section will be addressed, including $i)$ production, $ii)$ assembly, and $iii)$ integrating the photon detectors with all other prototype systems in a scaled down version of the full detector. 

A second important future direction is the development of realistic simulations to describe photon detection in LArTPCs.  Most simulations are currently in a preliminary stage of development.  These simulations are needed in order to determine, for instance, how much photon detector coverage is needed to address the physics questions these detectors are being designed to answer.  In addition, it is important to understand how the photon detector information will be integrated with the TPC information.  For LBNE this is a two-pronged analysis.  On the surface, the photon detectors are mostly needed for mitigating against cosmic and radioactive backgrounds.  Underground, the photon detectors are needed to mitigate against backgrounds for proton decay since there will be only very few events and all backgrounds must be identified with extremely high precision.  In addition, the photon detectors are necessary for the detection of the low energy neutrinos from core collapse supernovas.  Simulations are needed for both these physics analyses.  

Simulations are also needed for event reconstruction using photon detector information.  Reconstruction issues will be greatly aided by tuning the Monte Carlo simulations with detector data.  LArIAT is designed specifically to address the issue of the contribution of photon information into calorimetric reconstruction.  For MicroBooNE event reconstruction with photon detector information is actively being investigated with the detector currently being investigated.  One important motivation for the 35~ton prototype for LBNE is to address this very issue -- how to integrate photon information into reconstruction.  

\subsection{Synergies with the Dark Matter Community}

The LArTPC neutrino community has benefited from the developments in dark matter detector technology and event reconstruction.  Foils coated with a thin TPB film, as used in LArIAT, were pioneered by the dark matter community. The basic techniques of calorimetric energy reconstruction and pulse shape discrimination 
were also developed for dark matter experiments. 

The dark matter community can also benefit from the improvements in photon detection technology developed for the LArTPCs used in neutrino physics experiments.  The direct detection of dark matter WIMPs is being actively pursued with a number of different technologies.  One important technology being considered for the next generation of increasingly larger detectors uses liquid Xe as the active detector medium.  The scintillation light generated by WIMPs in LXe at 175~nm gives provides the information needed to reconstruct the WIMP events.  There are also several proposed dark matter experiments that use LAr.  In both cases, there are clearly strong synergies with the photon detection R\&D described here for neutrino physics with LArTPCs.  

In particular, dark matter experiments with liquid noble elements often use PMTs coated with waveshifter or PMTs behind waveshifting plates, like MicroBooNE.  In both, a VUV waveshifter that is more efficient at 175~nm  and 128~nm or one that is more cost-effective would certainly be valuable for dark matter experiments.  It is quite likely that future dark matter experiments will consider SiPMs as photodetectors.  Improved understanding of the role of nitrogen contamination in the attenuation and quenching of scintillation photons will benefit the searches for dark matter WIMPs.

\section{Calibration and Test Beams}
\label{sec:TestBeams}
\subsection{Calibration}

Calibration measurements performed before 2013 were not coordinated with a goal of implementing the results in future larger-scale neutrino and dark matter experiments.   The next generation of these measurements is intended to be part of a coherent plan that minimizes duplication of effort while addressing all of the relevant issues. This section describes both existing measurements and techniques developed before the need for a coherent strategy was recognized as well as future measurements that are being coordinated.

\subsubsection{Dark Matter Experiments}

The premise of experiments focused on the direct detection of dark matter is that dark matter is composed largely of a single heavy species of weakly interacting massive particles (WIMPs).  WIMPs will interact via low-energy elastic collisions with nuclei, on the order of keV.  As such, WIMP detectors are focused on measuring single, uncorrelated, low-energy nuclear recoils. The lower the energy threshold, the more sensitive the detector.

\begin{figure}
\centering
\includegraphics[width=3in]{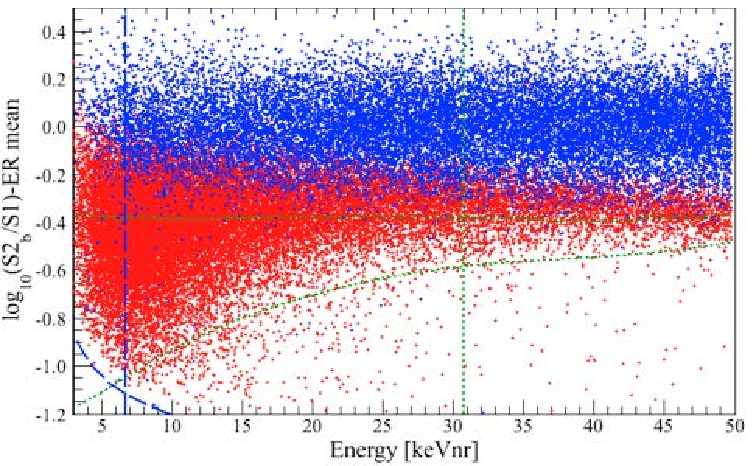}
\includegraphics[width=3in]{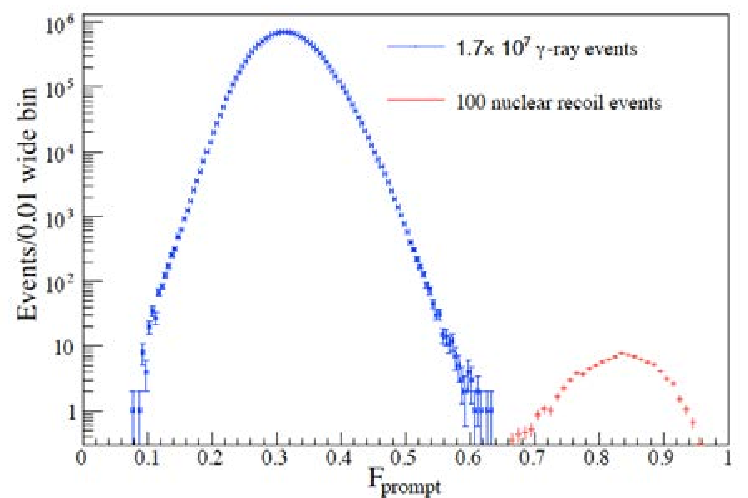}
\caption{\label{fig:DM_figs} In the top panel, differentiation of WIMP signals from EM backgrounds is achieved by taking advantage of the different ratio of deposited charge to produced scintillation light for each of the two categories. In the bottom panel, another method for distinguishing between heavy and light particles is the fraction of prompt light.}
\end{figure}

Detectors must be low background and have a sufficient response to detect low energy signals.  Noble liquid detectors take advantage of the natural low-background and the scintillation properties of the medium.  The ratio of charge deposition to produced scintillation light in such detectors allows WIMP signals to be separated from electromagnetic backgrounds such as betas and gammas from radioactive decay of isotopes in the detector materials, as seen in Fig.~\ref{fig:DM_figs}.  Heavy and light particles also populate different fractions of atomic states with different decay constants.  As such, the ratio of prompt to delayed light is also employed to discriminate between WIMP signatures and backgrounds.

The position and energy calibration strategy of such detectors is done with a combination of external and internal sources.  External sources include gamma and neutron sources.  The gamma sources give a well-defined electromagnetic energy and can be matched to simulations for a known source location. Neutron sources mimic the WIMP signal.  Internal sources include the introduction of krypton into the detection medium by way of a $^{83}$Rb source that is inline with the purification system.  $^{83}$Rb decays to $^{83}$Kr-m.  This metastable state decays via a 32 keV gamma-ray with a half-life of 1.83 hours, with a subsequent decay via a 9 keV gamma-ray with a 154.4 nanosecond half-life.  The nearly two-hour half-life of the first decay allows this source to be uniformly distributed in the detector.  The resultant $^{83}$Kr isotope is inert and stable.

Other strategies include looking for known surface radiation from radon and other contaminants.  In the case of argon, the probability of capture on argon from a fast neutron is negligible because of a large negative resonance above the thermalization energy.  Detectors employing argon are usually contained in an active water veto where a fast neutron that creates a signal in the inner volume can be detected upon its thermalization in the veto.

Energy calibration in dark matter liquid argon detectors is not in the same range as that for neutrino LArTPCs, however the use of scintillation light in both is an area where DM- and non-DM liquid argon experiments each can benefit from measurements and technological advances made by the other.

\subsubsection{Neutrino Experiments}

\paragraph{ArgoNeuT}

The ArgoNeuT experiment collected data in the NuMI neutrino beam line at Fermilab.  It has used beam-related muons, rather than neutrino interactions in the detector, to calibrate the detector.  The detector, with 480 channels instrumenting 175 liters of liquid argon, was able to use muons crossing the LArTPC matched with their tracks in the MINOS detector to measure the argon purity, electron drift velocity, energy deposited by minimally ionizing particles, and also to understand reconstruction and as-built detector geometry. A large sample of through-going muons from the beam was used for these measurements.

\begin{figure}
\centering
\includegraphics[width=3.in]{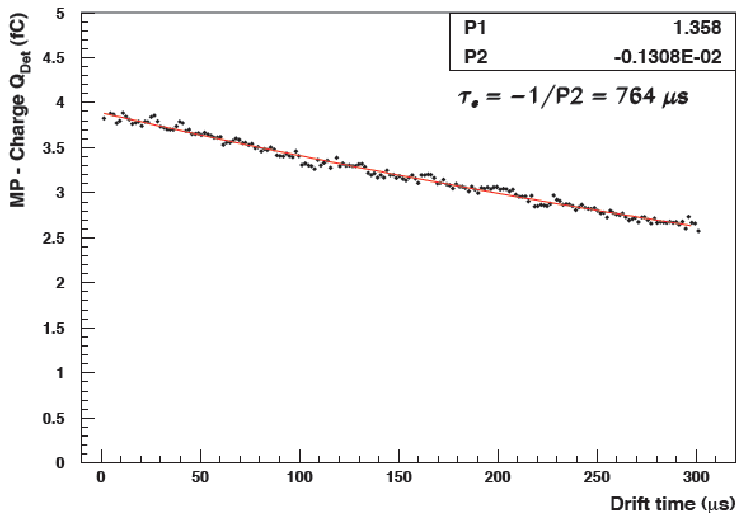}
\includegraphics[width=3.in]{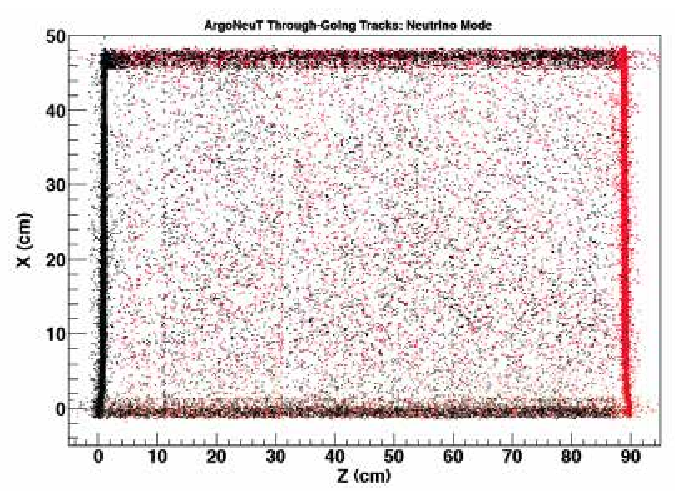}
\caption{\label{fig:argoneut_figs} In the top panel, a fit to the collected charge {\it{vs.}} drift time gives a measure of the electron lifetime. In the bottom panel, through-going muons were used to investigate possible electric field distortions in the detector.}
\end{figure}

One important value to monitor in LArTPCs is the electron lifetime, where fluctuations are indicative of an increase or decrease in purity of the argon.  The lifetime measurement is obtained by fitting an exponential to the distribution of charge as a function of the drift time, as shown in the top panel of Fig.~\ref{fig:argoneut_figs}. ArgoNeuT did this for the induction plane and also for the collection plane of the LArTPC, finding consistent results for the two independent measurements.

The drift velocity, which varies with electric field and with temperature of the argon, can also be measured using the distribution of hit times for long tracks in the detector. Dividing the maximum drift length of the tracks, with corrections for thermal contraction, by the maximum drift time gives a measurement of this value.

Measuring $dE/dx$ for the crossing muons gives a good calibration of the energy, since these are expected to be minimum ionizing particles (MIPs).  By fitting a histogram of deposited charge per unit track length and fitting it to a Landau+Gaussian, an average $dE/dx$ of $2.3\pm0.2$~MeV/cm was obtained.

Finally, muons can be used to image the detector, as seen in the bottom panel of Fig.~\ref{fig:argoneut_figs}.  This effective ``X-ray'' image can be used to reveal electric field distortions and to study the performance of reconstruction.

Most of these calibrations can also be achieved with a UV laser, but given that muons are free and readily available, they make an excellent calibration source for LArTPCs in the absence of a laser.

\paragraph{Laser Calibrations}

Lasers are a particularly attractive possibility for calibrations as they can create well-localized ionization tracks in the argon in space and time.  The prototype ARGONTUBE system at the University of Bern, has been employed to measure liquid argon purity and to determine electric field distortions.

\begin{figure}
\centering
\includegraphics[width=3.in]{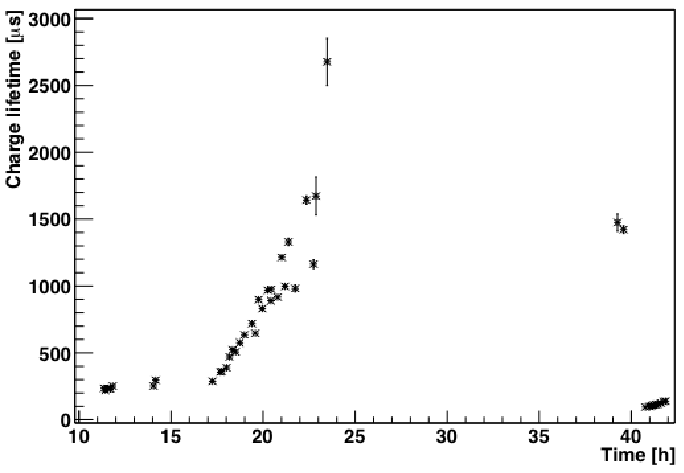}
\includegraphics[width=3.in]{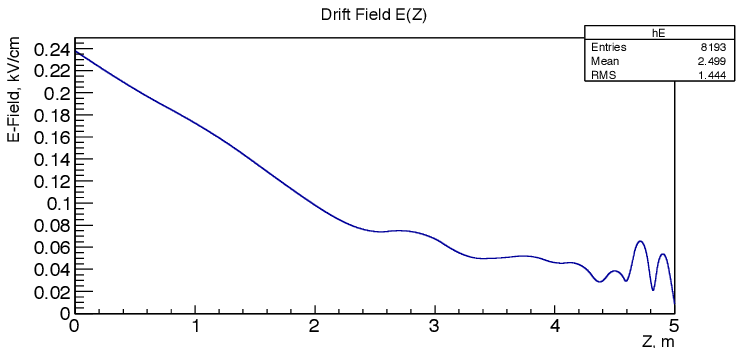}
\caption{\label{fig:laser_figs} In the top panel, electron lifetime is shown as a function of time. In the bottom panel, the laser is used to map imperfections in the drift field.}
\end{figure}

The Nd-YAG laser in this system ionizes liquid argon to create a track diagonally across the 20~cm width of the readout plane and across the 5~m drift length of the TPC.  The measured lifetime is shown as a function of time in the top panel of Fig.~\ref{fig:laser_figs}.  The increasing lifetime between 17 and 24 hours is the period when a recirculation system was engaged, followed by an overnight shutdown of the recirculation system.  The measurements starting again at 39 hours, at $\sim1.5$~ms lifetime, were taken after the experimenters returned to the lab.  The sudden drop just after 40 hours coincides with the event of refilling the detector with dirty argon, followed by restarting the recirculation system.

During the test, a capacitor in the HV distribution system broke, causing a non-uniform drift field.  Instead of immediately fixing the problem, the group realized they could take advantage of the situation and demonstrate the possibility of correcting field distortions using laser calibration data.  The distorted field is shown in the bottom panel of Fig.~\ref{fig:laser_figs}. Future shallow LArTPCs will have space-charge effects, a local accumulation of ions which could be as strong as the drift field, leading to distortion of the field. Corrections can be derived from knowledge of the laser path, allowing for extraction of the true drift field. 

Development of devices to steer the laser beam into various locations are also underway to allow for a detailed calibration of large volume detectors. The first such device will be demonstrated in MicroBooNE, and can be easily adapted to future detectors like LBNE and LBNO.

\subsection{Test Beams}

An active test-beam program to calibrate LArTPCs is being developed in the U.S., with complementary branches aiming to cover a wide range of measurements and calibrations that will be useful in upcoming near- and far-term experiments. The LArIAT (LArTPC In A Testbeam) program will study LArTPC response to charged particles in the $\sim200$~MeV to $\sim2$~GeV range, while the CAPTAIN (Cryogenic Apparatus for Precision Tests of Argon Interactions with Neutrinos) program will begin by studying low and medium energy neutrons in LArTPCs. Efforts to characterize LArTPC performance are underway in Japan as well, with the T32 test beam experiment that ran in a test beam at J-PARC in 2010, and an upgraded version of the same experiment that is preparing to run in late 2013. The LArIAT and CAPTAIN experiments are described in more detail below, while the T32 experiment is described in Section~\ref{sec:World}, which details on-going efforts in LArTPC R\&D outside of the US.

\subsubsection{LArIAT}
\label{sec:LArIATTB}

\begin{figure}
\begin{center}
\includegraphics[width=3.in]{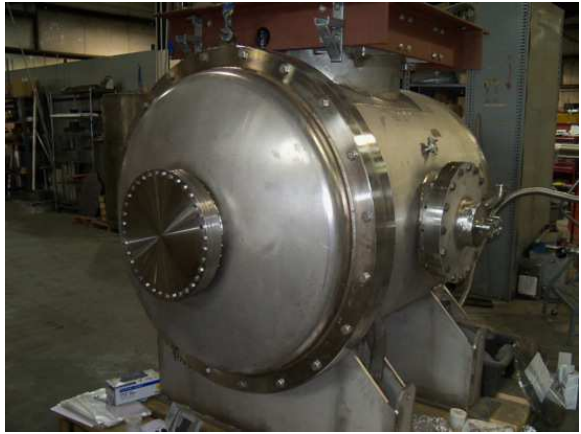}
\caption{\label{fig:lariat_detector} The LArIAT cryostat  showing upstream port for a Ti window, and side port for mounting photodetectors.}
\end{center}
\end{figure}

The LArIAT program will comprehensively characterize the performance of LArTPCs in the range of energies important for experiments such as MicroBooNE and LBNE.  The program is divided into two phases.  The first makes use of the existing ArgoNeuT detector, with modifications to make it suitable for collecting data in a charged particle beam instead of a neutrino beam, shown in Fig.~\ref{fig:lariat_detector}.  The second phase will collect data in the same beamline, but with a larger volume TPC that will contain a large fraction of the energy and extent of EM and hadronic showers.

Both phases of the experiment will be hosted in the Fermilab Test Beam Facility, which will have a dedicated liquid argon calibration and R\&D area with a highly configurable secondary or tertiary beam line created from the tunable primary proton beam.

In the first phase of LArIAT, the experimental program will include: $i)$ experimental measurement and proof of $e/\gamma$ separation in LArTPCs, $ii)$ measurement of the effects of charge recombination along ionization tracks for a range of particles and energies, $iii)$ optimization of particle identification methods using $dE/dx$ and residual range for single tracks, and $iv)$ development of charge sign determination criteria without a magnetic field.

\begin{figure}
\centering
\includegraphics[width=3.in]{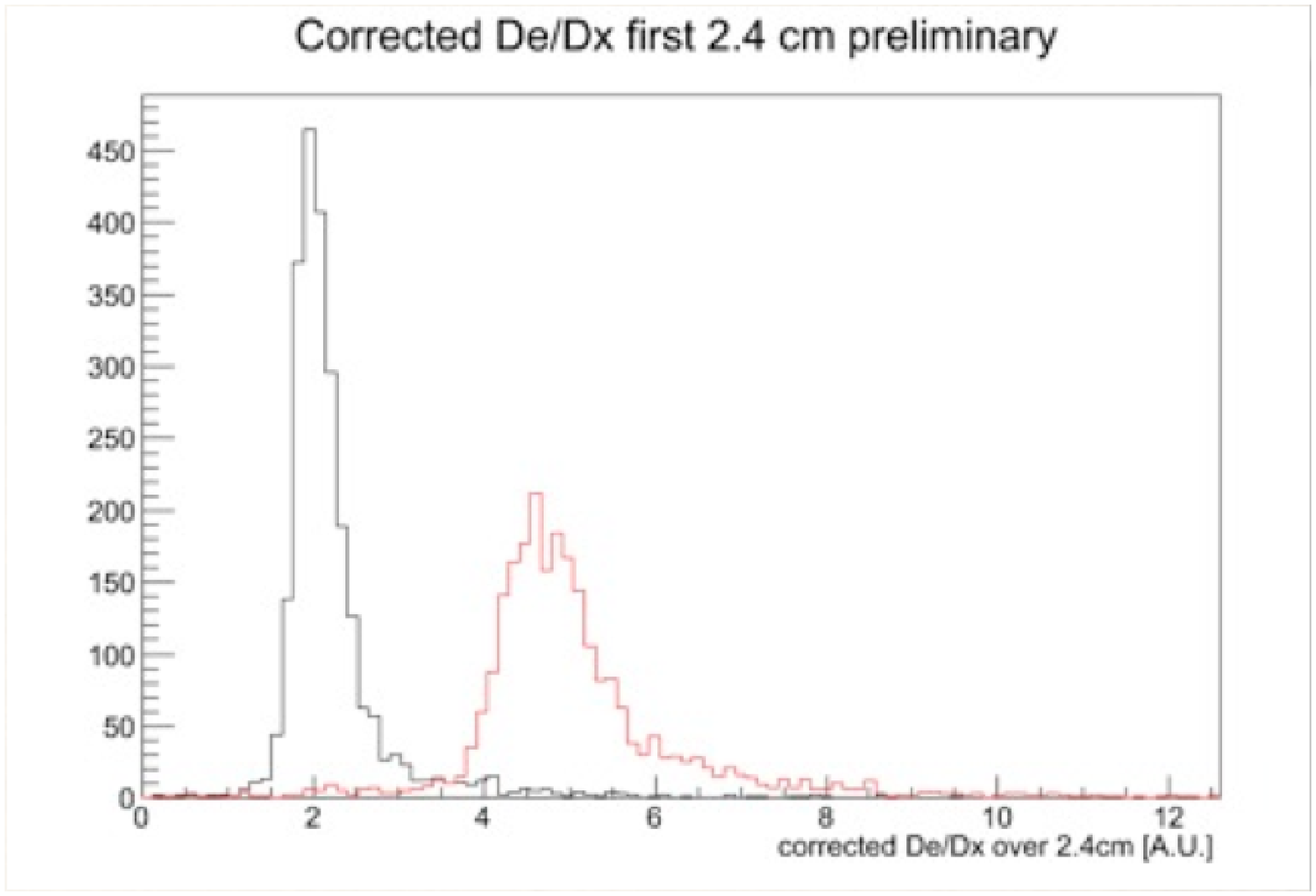}
\includegraphics[width=3.in]{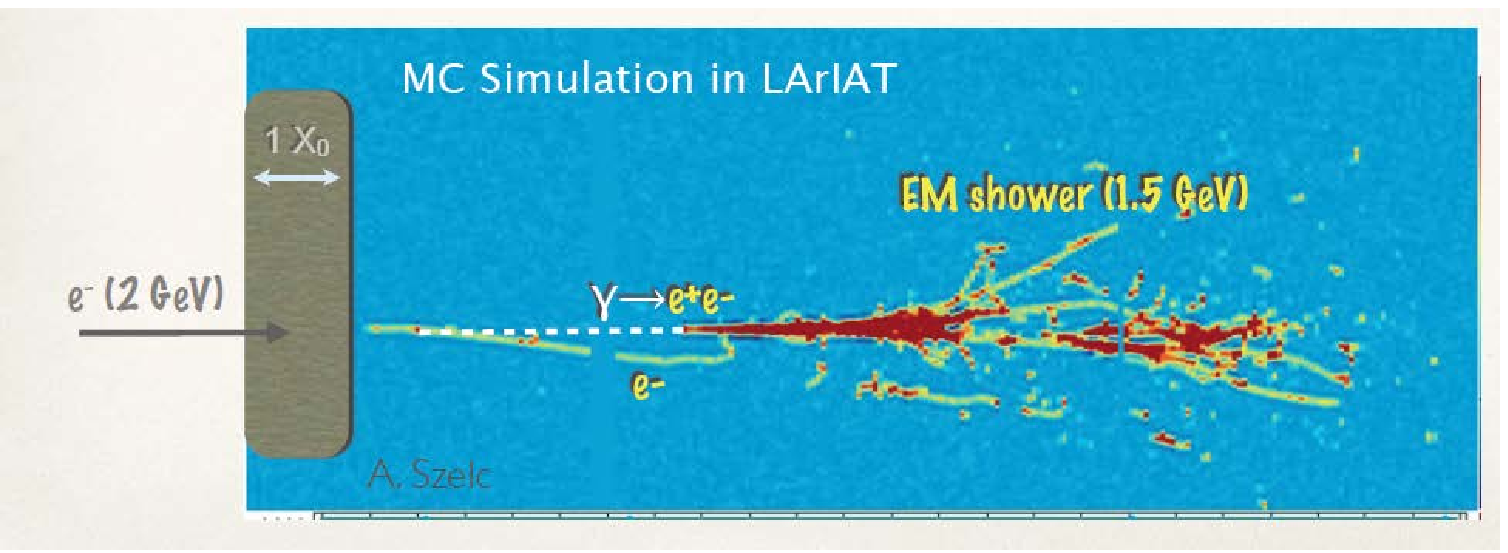}
\caption{\label{fig:lariat_Egammasep} Top panel: Electron-vs. photon-initiated shower separation by dE/dx (electrons in black,photons in red) for Monte Carlo simulation. Bottom panel: Simulated event with photon-initiated shower from bremsstrahlung of the incident electron.}
\end{figure}

Separation of electron-initiated from photon-initiated showers is critical for the upcoming MicroBooNE experiment, which will address the question of whether the low-energy excess seen in MiniBooNE arose from electrons, an indication of new physics, possibly sterile neutrinos, or from photons, an indication of an unaccounted background, possibly due to a previously unmeasured process. The first few centimeters of an EM shower in a LArTPC are the most important for determining the shower's origin; in these first few centimeters, a photon-initiated shower will leave double the ionization of an electron-initiated shower, since the photon only becomes visible when it converts to an $e^{+}e^{-}$ pair. A simulation of expected $dE/dx$ for the first 2.4~cm of the two types of showers is shown in Fig.~\ref{fig:lariat_Egammasep}. LArIAT will provide the first direct experimental measurement and verification of this capability of LArTPCs.

With the large sample of charged particles collected in the test beam, LArIAT will also determine the calibration from measured charge to incident particle energy. This requires precise measurements of electron-ion recombination in the argon for a range of energy deposition and different electric field values in the $\sim300-1000$~V/cm range, and at different track-to-electric field angles. The ideal tool for studying these is a test beam with particles that penetrate and slow down to stop in the LArTPC. The same sample of stopping particles will provide an excellent test bed for further development and refinement of particle identification algorithms.

The possibility of determining the sign of charged particles in a LArTPC has not been explored, but would be a powerful tool if it is demonstrated. Even in the absence of a magnetic field, it may be possible to statistically separate positively and negatively charged muons based on topological differences between muon capture and muon decay: positively charged muons always decay, while negatively charged muons capture 75\% of the time and decay only 25\% of the time. LArIAT will make a systematic study of capture in argon and LArTPC sign-selection capabilities.

The first phase of LArIAT will also be dedicated to small-scale R\&D. As described in \S~\ref{sec:LArIATPhot} a PMT-based light collection system installed in the cryostat, viewing the active volume through the wire planes, aims to use methods adopted from dark matter searches to augment the calorimetric reconstruction and particle identification in neutrino experiments. One goal is to test the use of pulse shape discrimination as a possibility for improving standard particle identification methods.

The second phase of LArIAT will extend the reach of the experimental program by addressing another important category of LArTPC calibration, calorimetry. This phase will employ a 2~m $\times$ 2~m $\times$ 3~m LArTPC to provide a conversion from measured energy to incident energy for more complicated event types such as electromagnetic and hadronic showers. The goal is to understand the size and features of each type of shower.  Each of these two shower types has its own set of challenges.  For example, in an EM shower $\sim$30\% of the incident energy goes into soft electrons; developing strategies to precisely reconstruct the incident energy are crucial to future experiments.  Hadronic showers are significantly more complicated, involving a greater number of particle types and developing over a much larger spatial extent, which makes the larger volume LArTPC crucial for these studies. In addition, the phase two detector will serve as a testing ground for liquid argon subsystems under development such as cold electronics and new wire plane designs.

The LArIAT program will collect data in the Fermilab test beam starting in FY14.

\subsubsection{CAPTAIN}
\label{sec:CAPTAINTB}
\begin{figure}
\begin{center}
\includegraphics[width=3.in]{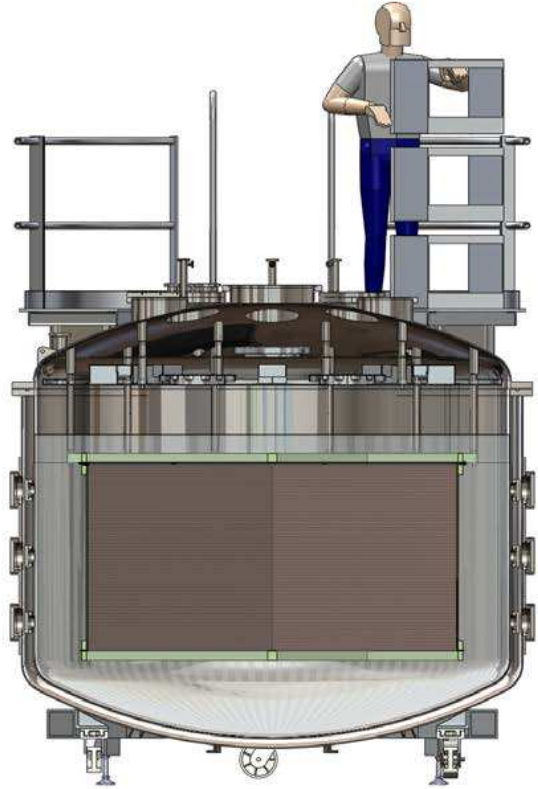}
\end{center}
\caption{\label{fig:captain_setup} The CAPTAIN detector setup.}
\end{figure}

The CAPTAIN experiment is being developed under the auspices of a Laboratory Directed Research and Development (LDRD) project at Los Alamos National Laboratory (LANL).  The detector, shown in Fig.~\ref{fig:captain_setup}, is housed in an 8,000 liter vacuum insulated cryostat.  Five tons of liquid argon are instrumented with a 2,000 channel TPC and a photon detection system.  The TPC detector for this assembly has a wire length of 2~meters and a drift distance of 1~meter. This TPC is designed as a hexagon and it will make use of a high voltage feed-through from UCLA that will operate at 50 kilovolts. The cryostat has ports that can hold optical windows for laser calibration.

The program of measurements being developed is designed to address several physics questions important for LBNE's long-baseline neutrino oscillation program, atmospheric neutrino program and burst supernova neutrino program.

In the commissioning phase, laser calibration, cosmic-ray, and possible source calibration measurements will be made.  An external muon tracker has been assembled for the cosmic-ray data.  Detailed studies of positive and negative muons will be made to determine the optimal strategies for separating them in the non-magnetized LBNE far detector.  Separating positive and negative muons will be particularly useful in LBNE's anti-neutrino run because there is a large neutrino contamination in the beam, and for the analysis of atmospheric neutrinos which contain both neutrinos and anti-neutrinos.  Preliminary simulations have been carried out on neutron signatures.  Potential gamma and neutron signatures will be studied in situ. In addition, the steady-state low-energy background to supernova neutrinos will be studied.  The laser calibration system will be employed to calibrate the detector response and to develop the system for LBNE.

\begin{figure}
\centering
\includegraphics[width=3.in]{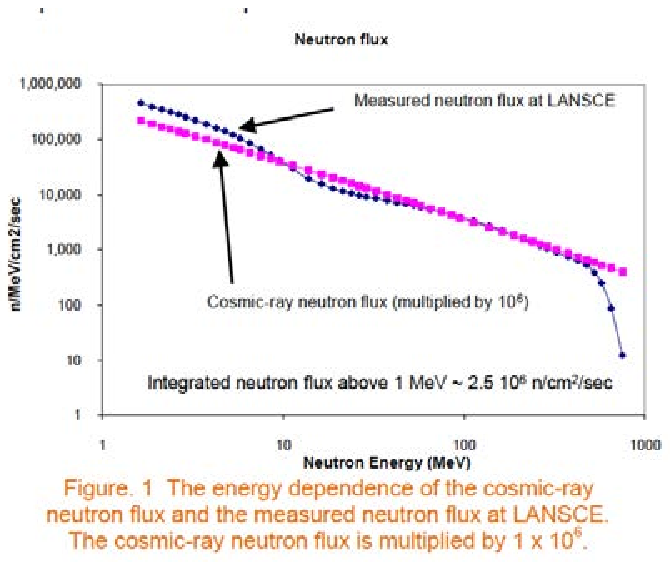}
\includegraphics[width=3.in]{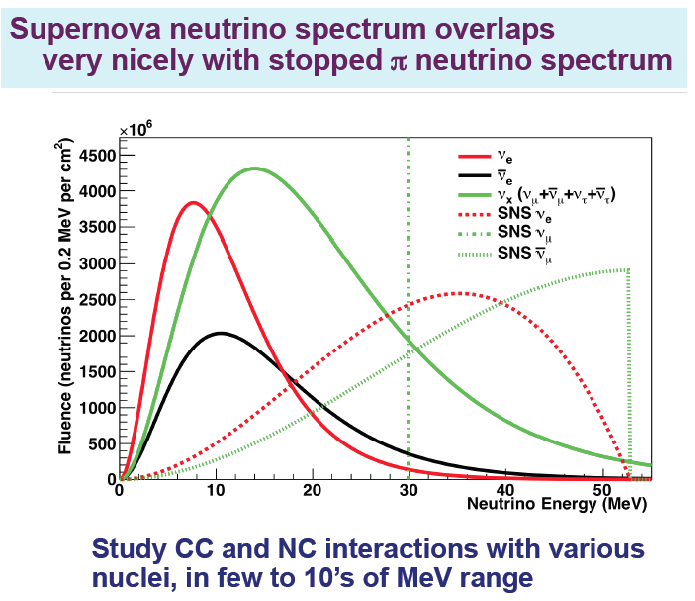}
\caption{\label{fig:captain_figs} The top panel shows energy range and flux of neutrons at LANSCE. In the bottom panel, the energy spectrum of neutrinos from the SNS stopped pion neutrino source is in the same range as that of supernova neutrinos. }
\end{figure}

The next phase will involve data taking in the Los Alamos Neutron Science Center (LANSCE) high-energy neutron beam.  At LANSCE, there are two classes of goals: the first supports lower energy neutrino analysis, namely supernova neutrinos, and the second supports higher energy LBNE neutrino analysis, namely beam and atmospheric neutrinos.  In the low energy range, CAPTAIN will look at the production by neutrons of isotopes that will constitute the background to supernova neutrino interactions.  Time-of-flight techniques allow the measurement of neutron energy-dependent production cross-sections in each case.  The neutron spectrum is shown in the top panel of Fig.~\ref{fig:captain_figs}. At higher energies, CAPTAIN will make studies of single neutron interactions in a kinetic energy range of several hundred MeV to search for events that could mimic electron neutrino appearance and to develop strategies to constrain the hadronic energy lost to neutrons in hadronic showers from neutrino interactions.

Following neutron running at LANSCE, the portable detector will be moved to a neutrino beamline.  Currently, the most scientifically interesting options appear to be running in an on-axis position in the NUMI beamline or running in the stopped pion neutrino source created in the production of neutrons at the Spallation Neutron Source at Oak Ridge National Laboratory.  NuMI will be running in the medium energy tune for the NO$\nu$A experiment during this period.  The spectrum of neutrino energies in the on-axis position is complementary to MicroBooNE and results from the two experiments together will provide a program of measurements in the neutrino energy regime important for LBNE.  MicroBooNE will thoroughly explore quasi-elastic scattering and delta-resonance production.  CAPTAIN in an on-axis position in the NUMI beamline would explore higher-order resonances and the deep-inelastic scattering regime.  CAPTAIN deployed at SNS would make cross-section measurements of neutrinos on argon in an energy regime overlapping that from supernovae, as shown in the bottom panel of Fig.~\ref{fig:captain_figs}. This would be the first measurement of the argon cross-section in this energy range, and also the first demonstration of identification and reconstruction of charged-current neutrino reactions at supernova energies in an LArTPC.

The current schedule involves commissioning during FY14 with neutron running in FY15 and neutrino running beginning in FY16.

\subsubsection{Test Beams Summary}

The U.S. HEP community is developing a coherent plan to address all the pertinent questions regarding LArTPC performance and calibration. CAPTAIN will operate in the LANSCE neutron beam to study neutron interactions in argon. There also exist possibilities for operating CAPTAIN in the SNS stopped pion neutrino beam to study low energy neutrino interactions in the range relevant to supernova burst neutrinos, and in the NuMI beam to study higher energy neutrino interactions. The LArIAT test beam program will operate in parallel with CAPTAIN, but with a complementary set of goals. LArIAT will characterize the performance of LArTPCs using the charged particle test beam at Fermilab, studying particles in the energy range relevant to MicroBooNE and LBNE. In addition, LArIAT will make the first direct experimental measurement of LArTPC efficiency for distinguishing electron-initiated showers from photon-initiated showers. In LArIAT's second phase, the larger volume LArTPC will contain hadronic and EM showers, allowing development of calorimetric reconstruction and measurements of energy resolution. The menu of calibration measurements mentioned above is also being pursued outside of the U.S., specifically in Japan with the T49 test beam experiment. Verification of results among all the programs will give confidence to the planned future program of neutrino physics using LArTPCs.

\section{Software}
\label{sec:Software}
Existing liquid argon time projection chamber experiments such as ArgoNeuT and ICARUS, and upcoming ones such as MicroBooNE and LBNE, make exciting neutrino physics measurements and discoveries possible. The simulation and reconstruction of the relevant signal and background processes, however, poses significant challenges, largely due to the vast amount of precise tracking information provided by these detectors.  Efforts are underway to make use of the latest techniques to improve the detail and accuracy of the simulation, as well as the speed and accuracy of the reconstruction.

The design of liquid argon TPC detectors, the estimates of their performance, and the extraction of physics results from them requires detailed simulation and reconstruction.  The current state of the art in detector simulation is the Geant4 package~\cite{geant4}.  Its object-oriented structure provides for interfaces with modern applications written in C++, such as the LArSoft~\cite{larsoft} package for simulation, reconstruction and analysis of LArTPC data.  Reconstruction algorithms must contend with the large amounts of raw data produced by liquid argon TPC's and produce meaningful physics objects and associated uncertainties.  Characterization of the performance of the reconstruction algorithms to predict signal and background acceptances and particle identification and mis-identification rates is done with the full Monte Carlo simulation, backed up by measurements in data.

Several groups are actively participating in the development of LArSoft.  These groups often contribute to LArSoft in order to further at least one of ArgoNeuT, MicroBooNE, LBNE or LArIAT.  That is, LArSoft is designed to be used by multiple experiments and people working on multiple LArTPC experiments can contribute to a single body of code in LArSoft.  The ArgoNeuT collaboration has lead the software development effort in order to analyze their data, producing several papers in the process~\cite{argopapers}.  As the other experiments become more mature, they are adding significant contributions to the software too.  The contributions described below come from Duke University (Li, Scholberg), Fermilab (Baller, Carls, Greenlee, Junk, Kirby, Lockwitz, Rebel), Louisiana State University (Insler, Kutter), MIT (Jones), Otterbein College (Tagg), the University of South Carolina (Alion, Kim, Mishra), Syracuse University (Asaadi, Soderberg), and Yale University (Church, Palamara, Partyka, Szelc).

\subsection{Ionization Simulation}

Activity in LArTPC simulation currently focuses on incorporating LBNE into the simulation and geometry description tools and developed in LArSoft and this work is being done by the South Carolina and Fermilab groups.  The two detectors currently envisioned for LBNE are a 10 kT surface far detector and a 35T prototype detector under construction at Fermilab.  Digitization for LBNE is being developed by the LSU group and photon detection for LBNE is being developed by the Duke group.  Both the Yale and LSU groups are collaborating on the noise simulation.  

Both the far detector and the 35T prototype have more complicated geometries than MicroBooNE and ArgoNeut.  Specifically, multiple drift volumes exist in the LBNE geometries, while the other two detectors have a single drift volume.  The drift volumes in LBNE have oppositely directed drift fields, as charge drifts towards a centrally located Anode Plane Assembly (APA) from two directions.  The LBNE geometries are further complicated by the fact that the induction-plane wires wrap around the front and back sides of the APA's.  This feature has the consequence that DAQ channels correspond to multiple wire segments, which view argon in different drift volumes, whose signals are added together, creating ambiguity in the interpretation of the data.

A realistic noise simulation is being developed for MicroBooNE.  An arbitrary spectral function can be put into the LArSoft simulation tools.  Random noise is generated from these spectra using a FFT, with random phases for each frequency component.  The same noise spectrum is used to construct a Weiner filter used to filter out noise, although a signal spectrum must be assumed.  The signal and noise frequency response functions are input to the Weiner filter.  As the variety of expected signal shapes is large -- tracks may be incident at any angle and particles of different types yield signals that have showers, delta rays, or simply have different ionization densities, it is important to keep the noise as small as possible so that the noise filter removes as little of the signal as possible.

Future directions include improving the sophistication of the material modeling, such as including electronics and cables.  Radiological modeling also needs to be added.  Photon detector modeling is being worked on.  The display of data from LArTPCs needs additional features to improve its utility for all experiments and the Otterbein group is working on those improvements.

\subsection{Ionization Reconstruction}

The LArTPC reconstruction effort consists of several parallelizable stages.  In LArSoft, the raw data are first passed through a calibrated deconvolution algorithm, which filters noise and corrects for the electronics response and the effect of the drift field response, such as the bipolar signals on induction wires, to produce the best estimates of the drift charge and time.  Hits are then found in the deconvoluted data with a Gaussian fitter.  These hits are then clustered, and the clusters are input to track reconstruction and shower reconstruction algorithms.  The primary vertex must be identified in neutrino scattering events.  Particle assignments must be made to the tracks and showers in order to extract physics results from them, and a global interpretation of the event collected together so that event selection requirements can be applied, and, if successful, the event can be entered into a histogram of reconstructed neutrino energy.

The Syracuse group has created a hit finding algorithm that fits Gaussians to the deconvoluted data and produces an ntuple with three entries -- position, width, and area.  The hit width is sensitive to the angle of the particle's trajectory with respect to the drift direction, and the hit area is sensitive to the particle type.

Hits from two or three views with similar arrival times are combined into $(x,y,z)$ space points.  Many combinations of hits in different views give rise to a high fake rate of space points, especially for tracks that move parallel to the readout wire planes.  Clustering algorithms collect hits together ideally to separate hits that originate from one particle from those made by other particles. Several clustering algorithms exist within LArSoft.  A density based scanning algorithm~\cite{dbscan} implemented by the Yale group  can be used to cluster hits with an average density above a threshold.  It is not aware of physical event features, such as vertices or kinks and as such needs additional information to separate hits associated by density into physical clusters.  A Hough line finder is available that can find straight track segments in one view at a time, for either isolated track finding or for finding substructure in showers.  The Fermilab  group has implemented a clustering algorithm, called ``Fuzzy Clustering''~\cite{fuzzy}, which merges nearby Hough lines in space and angle, and then merges nearby isolated hits.  This algorithm does account for physical features in the event to produce clusters that are associated with single particles.

An event reconstruction tool called PANDORA~\cite{pandora}, initially developed for linear collider detector simulation physics studies, is being adapted for use with LArTPCs in LArSoft.  PANDORA forms clusters of hits in 2D using topological associations such as proximity and pointing. PANDORA makes end-to-end merges, creating extended clusters, and identifies the ``spine'' of the event.  A primary vertex is identified, and particles emitted from that vertex are identified using length-growing algorithms to grow tracks and branch-growing algorithms to grow showers.  Topological associations then are used to add remaining clusters.  PANDORA has been shown by LBNE collaborators at Cambridge to work with LArSoft, but it is not yet fully integrated into LArSoft.

Tracks are reconstructed in LArSoft with a variety of algorithms: merged Hough lines, 3D Kalman filtering~\cite{kalman} using hits or space points, and 3D seed tracks.  The Yale group and Fermilab group have been working on track reconstruction algorithms using Kalman filters.  The deliverables are the start and end points in 3D, trajectories, charge deposition $dQ/dx$ along the track, the track direction, and a momentum estimate.  The momentum can be estimated from the range in an unmagnetized liquid argon detector for a stopping track, by the bend observed if the detector is magnetized, or by multiple scattering for a track that exits an unmagnetized detector.

The target deliverables of shower reconstruction are the shower vertex and axis in 3D, the total shower energy, and electron/photon separation, based on $dE/dx$ in the initial part of the shower.  The Yale group (Szelc) is working to develop a shower reconstruction that starts from clusters, and builds a hypothesis of the shower parameters in 3D by combining information from the 2D shower reconstruction in the separate views.  Axis reconstruction can be pulled off by outlying hits, and the vertex reconstruction can be confused by backward-emitted photons.  A current algorithm looks for track-like substructure to help identify the shower vertex.

Calorimetry of showers is important to reconstruct the total energy of electromagnetic showers, which is an important ingredient in reconstructing the signal in $\nu_e$ appearance analyses, as well as the rejection of $\pi^0$ backgrounds.  It is therefore necessary to use the collected charge $Q$ to best estimate the deposited energy $E$.  Effects from electron lifetime, depending on the drift distance, and the particle's angle with respect to the drift field are included. Of more concern is the modeling of ionization in an environment dense in tracks.  Birks' formula~\cite{birks} applies to charge depositions from a single charged particle.  In an electromagnetic shower, this is not the case, and so the ionization model may need to be modified. One model currently in use in noble liquid dark-matter detectors is NEST~\cite{nest}.  NEST introduces a statistical anti-correlation between the ionization charge yield and the photon yield for each ionization event along a particle's path.

Optical detectors are crucial in separating cosmic-ray events, which are a large background for TPC's on the surface, from neutrino interactions.  Cosmic rays can also be rejected by identifying tracks that are out of the spill time or that enter or exit the fiducial volume, especially from the top or the bottom.

\subsection{Photon Simulation and Reconstruction}

The ability to use light collected from neutrino interactions in LArTPCs, in addition to the ionization, is one of the attractive features of this detector technology.  Liquid argon is a very bright scintillator, emitting thousands of photons per MeV of deposited energy.  This scintillation light is very deep in the UV spectrum, at 128~nm wavelength, and argon is transparent to this light.  The timing of this scintillation light can be used to discriminate amongst beam and non-beam sources, as well as providing geometrical information on the origin of any light.  Current efforts to understand photon simulation and reconstruction in neutrino interactions are driven by the MicroBooNE experiment, which will employ 30~eight-inch Hamamatsu photomultiplier tubes.

The simulation of photons in LArTPCs is currently approached using two different modes.  In the first mode, a full simulation of photons produced in argon and propagated to the light collection system is utilized.  In this full simulation, a complete suite of physics processes, including scintillation light production, Rayleigh scattering, and Cherenkov production, is available in the Geant4 ``physics list" used during the stepping of the photons through the detector volume.  Each process in the physics list has adjustable parameters that can be controlled by the user.  The full simulation also contains information about the geometry of the detector environment, which is important for understanding the optimal layout of the light collection system.  The light collection system itself can be adjusted, depending on whether the experiment utilizes PMTs or some other form of light detector.  The full simulation has the best knowledge of the behavior of light created in LArTPCs, but has the drawback of being very computationally demanding and hence slow in execution.

The other mode of light simulation is dubbed the Fast Photon Table Sampling Simulation, which builds upon the full simulation.  In the fast simulation, a library has been created using the full simulation that contains the response of the light collection system to point sources of light that originate from a known coordinate in the detector volume.  The library allows a quick determination of how many photons from a given location in the detector will reach each optical detector in the experiment.  The fast simulation, relying on the pre-generated response library, is used to quickly arrive at the final optical system signal given some input distribution of light.

Once the light signal from the simulation or from data is in hand it needs to be reconstructed so the information can be utilized in analysis.  The primary use of the light information is a precise determination of the $t_0$ of the neutrino interaction, using the nanosecond resolution of typical optical detectors.  This same resolution also allows light originating from interactions occurring outside the beam spill window, such as cosmic ray tracks crossing the detector,  to be identified and subtracted.  The signal from each optical detector in the experiment is searched for hits, and if a grouping of optical detectors all have hits at the same time this is called a ``flash". Flashes are correlated to tracks observed through their ionization in the detector volume, and any out-of-time tracks can be identified and subtracted.


\subsection{Analysis Tools}

In addition to creating software tools that are robust enough to simulate and reconstruct neutrino interactions in LArTPCs, ideas are also being developed for physics analyses to perform with these tools. Among the topics of interest is the impact of nuclear effects on neutrino-argon interactions.  Particles created in neutrino-argon interactions must propagate through a dense nuclear environment before emerging as visible particles in the detector.  During this transit through the struck argon nucleus, the particles from the neutrino interaction may emerge without interacting with the nucleus, or they may be absorbed by the nucleus, or they may end up dragging  additional particles out of the nucleus that were not produced in the instigating neutrino interaction.

The implication of this rearranging of the final state makeup due to the nuclear environment is that the traditional classification of neutrino events, \textit{e.g.} Quasi-Elastic, Resonant, Deep Inelastic Scattering, is ambiguous and could lead to the misinterpretation of the underlying interaction process.  The ArgoNeuT collaboration, which already has a sample of LArTPC neutrino interactions, has begun to develop analyses that identify events based on their final-state topology, \textit{e.g.} $\mu^{\pm}$+ $N$ protons + $M$ pions, as opposed to the traditional categories~\cite{ornella}.  The Yale and Fermilab groups have worked to identify final-state particles in the interactions by measuring the calorimetric energy profile of candidate tracks.  ArgoNeuT is able to identify protons with kinetic energies $\geq$21 MeV, and this low threshold allows studies of the low-energy particles being produced in neutrino interactions on argon.  Early comparisons of data with different MC generators such as GENIE~\cite{genie} and GiBUU~\cite{gibuu} are promising, although discrepancies between the measured and predicted multiplicities and spectra of final-state particles are beginning to show.  ArgoNeuT collaborators are in contact with the authors of the different MC generators to incorporate some of the ideas being extracted from these studies.

\section{World Wide R\&D Efforts}
\label{sec:World}

\subsection{Japanese Efforts}

The completed T32 experiment at the J-PARC K1.1BR test beam used a roughly ArgoNeuT-sized TPC inside a 250-liter cryostat to collect a large sample of kaon, pion, electron, and proton events at momenta from about $500 - 800$~MeV/c. The readout was very coarse for this detector, consisting of a single plane of 76 1~cm wide readout strips all oriented vertically with respect to the beam direction. The coarse readout limited the amount of useful calibration that could be done with the dataset collected in 2010. 
\begin{figure}
\centering
\includegraphics[width=2.5in]{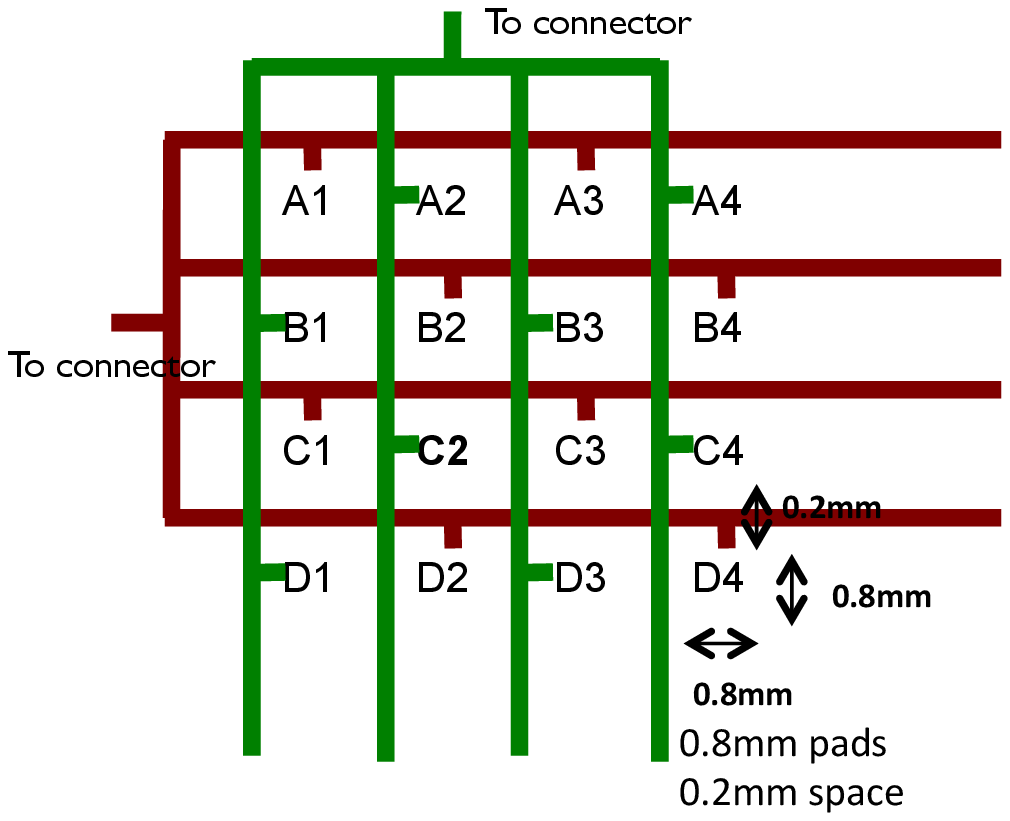}
\includegraphics[width=2.5in]{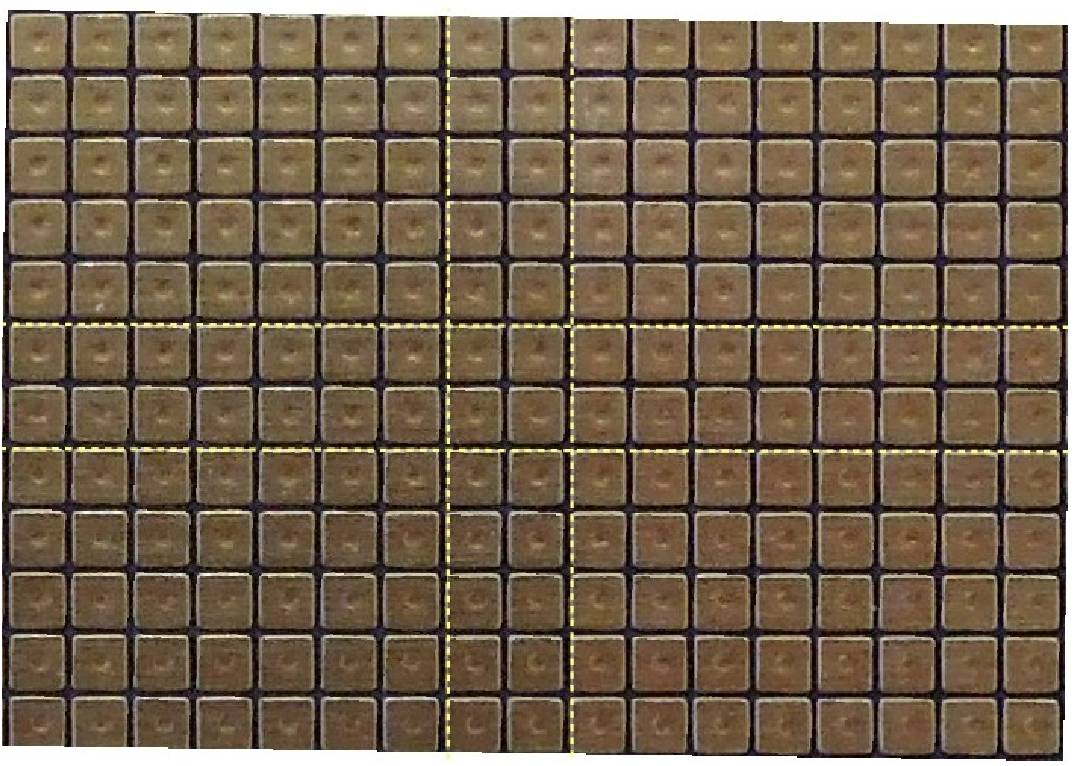}
\caption{\label{fig:T49_readout} The left panel shows the scheme for achieving 2-dimensional readout on a single readout plane. The right panel is a zoomed photograph of the actual hardware. }
\end{figure}

An upgraded system for T49, to operate in late 2013, will use the same cryostat and TPC cage, but with an improved liquid argon purification system and a newly-developed 2-dimensional readout using pads with 1~mm pitch connected to about 400 electronic channels, shown in Fig.~\ref{fig:T49_readout}. This represents an important step in the development of LArTPCs: demonstration of good performance of this pad readout scheme could eliminate the need for the delicate operation of winding and installing long wires, as is done in the current generation of LArTPCs.  Aside from the development and testing of the new style readout, T49 will also characterize and calibrate the LArTPC response for various test beams important for neutrino physics and proton decay experiments. The goals here are therefore very similar to those of LArIAT, including the investigation of particle ID performance, gamma/electron separation, characterization of event topologies, study of hadron transportation, and measurement of the energy resolution for EM and hadronic showers.

\subsection{European Efforts}

\subsubsection{ICARUS}
A major european effort in developing LArTPC detectors comes from the ICARUS program which was the first to demostrate the feasibility of this technology at the kiloton scale. The ICARUS T600 detector is the largest liquid argon imaging detector built to date, with over 500 tons of active mass~\cite{Amerio:2004ze}. It is located at the INFN Gran Sasso  underground Laboratory (LNGS) with an overburden of 1400 meters of rock. The design and assembly of the ICARUS T600 relied on industrial support and represents the application of concepts matured in laboratory tests. The T600 detector has been smoothly and safely run from May, 2010 to June, 2013, taking data from the CNGS neutrino beam and cosmic rays.  The concentration of impurities during data taking was consistently lower than 0.1~ppb and the detector was very stable with high live-time~\cite{Rubbia:2011ft}.  The  maximum achieved free electron lifetime during the running was about 6 ms with a 1.5 m drift. 

\begin{figure}
\begin{center}
\includegraphics[width=3.5in]{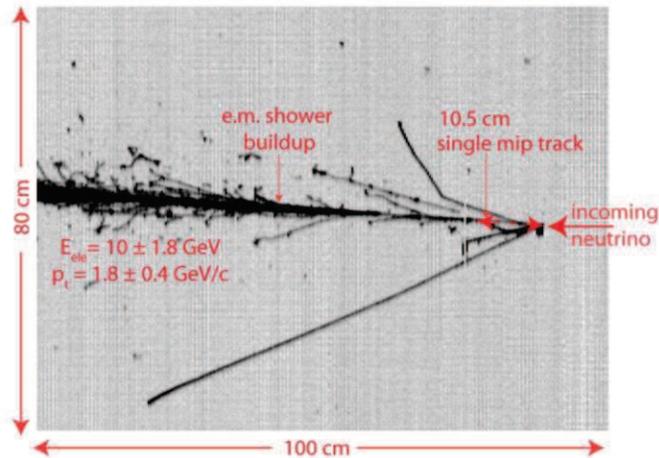}
\caption{\label{fig:ICARUSevent}Experimental picture of a $\nu_e$CC candidate with a clearly identifed electron signature. 
The candidate interaction has a total energy of 11.5$\pm$1.8 GeV and a transverse electron momentum of 1.8$\pm$0.4 GeV/c.}
\end{center}
\end{figure}

The ICARUS T600 Collaboration has produced several results from the analysis of CNGS neutrinos, including refutation of the superluminal neutrino claim~\cite{sln1}~\cite{sln2} and testing the LSND anomaly by searching for $\nu_e$ appearance in the CNGS $\nu_\mu$ beam~\cite{icalsnd1} \cite{icalsnd2}. The impressive spacial and calorimetric resolution of the detector, seen in Fig.~\ref{fig:ICARUSevent}, allowed for a clear separation of electrons and photons, which is crucial to search for a $\nu_e-$CC signature. The long term stability of ICARUS T600 detector has demonstrated that a LArTPC at large scale can be effectively used to in a neutrino oscillation experiment. The ICARUS collaboration is exploring the option of moving the T600 detector into a short-baseline neutrino beam along with another, new detector, to explore the origin of the LSND anomaly~\cite{icafnal}~\cite{icarusnessie}. The R\&D activities toward this goal have been officially recognized by CERN as the WA104 experiment. The proposed experiment could also be used to test at small scales, on the order of 0.1~--~1 kiloton, new solutions which could be adopted by future multikiloton LArTPC detectors like LBNE, such as potentially using magnetized LArTPCs and exploring new light collection and electronics technologies.

\subsubsection{Dual-Phase TPCs}
\begin{figure}
\centering
\includegraphics[width=3.5in]{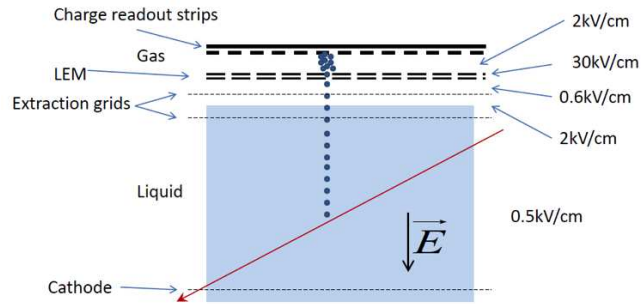}
\caption{\label{fig:dual_phase} Dual phase LArTPC operation.  The active electron gain element is the LEM, a large area electron multiplier.  The LEM thickness is about 1mm, and the gain is about 30.  The distances between grids and planes at the top of the figure is about 2~cm, and the liquid argon volume can be scaled to achieve the desired active mass.}
\end{figure}

A large part of the European efforts are devoted to developing dual phase detectors, with electron amplification produced by a plane of Large Electron Multipliers (LEM) in the gas phase above the active liquid volume. Figure~\ref{fig:dual_phase} is a diagram of the operational concept for such a LArTPC. The ultimate goal of these efforts is a 100kT scale dual phase Ar TPC as the far detector for a long-baseline neutrino observatory. The LAGUNA-LBNO collaboration has submitted a letter of interest to the CERN SPSC for 50~kT and 20~kT dual phase LArTPCs at the Pyhasalmi mine, with a 1400~km baseline from a proposed neutrino source at CERN.

This work is lead by the ETH group, which has successfully constructed a 200~liter prototype for a dual phase detector with a $40 \times 76$~cm$^2$ LEM readout surface and a 60~cm electron drift in LAr. The LEM readout has an effective gain of 14, and a signal to noise ratio greater than 30 for minimum ionizing particles~\cite{Badertscher:2012dq}.  Two charge collection planes following the LEM provide two independent, projective readout views, each with 3~mm pitch.  The drift field is produced by a Greinacher (Cockroft-Walton) voltage multiplier operated in the liquid argon, to avoid difficulties in constructing and operating the very high voltage feedthrough required for long drift distances.  This effort is summarized in more detail in \S~\ref{sec:HV} above. In addition, the University of Liverpool and Saclay are investigating the use of MicroMegas, as an alternate to the LEMs, to provide much higher gain in the gas. The next step planned in this program is the construction of a $6 \times 6 \times 6$~m$^3$ prototype, with a 36 ~m$^2$ LEM readout surface, to be constructed and operated in a test beam at CERN.

\section{Summary}
\label{sec:Summary}
Rapid progress is being made in each of the areas listed above.  The various efforts across the US are complimentary and collaborative as seen in Table~\ref{table:whoswho} which lists the people who organized the workshop sessions, those working in each of these areas, their institutions and their experimental affiliation.

\begin{table*}
\caption{\label{table:whoswho}  Participants in each of the primary areas of LArTPC R\&D. For the Calibration and Test Beams category, only the spokespeople of the CAPTAIN and LArIAT collaborations are listed.  Similarly for Software, the LArSoft project manager is listed.  Naturally the collaborations include many more people than can be listed here.} 
\centering
\resizebox{\textwidth}{!}{\begin{tabular}{|c|c|c|c|c|} 

\hline \hline

 R\&D Category (Organizer)                                     &  Institution  & Contact           & Experimental Affiliation\\ \hline 
 Argon Purity  (Pordes)                                             & Fermilab    & Pordes, Rebel & Generic R\&D \\ \hline
 Cryogenics    (Cavanna)                                          & Fermilab    & Rebel               & Generic R\&D \\ 
                                                                                  & Fermilab     & Hahn               & LBNE \\ \hline
TPC and High Voltage                                               & UCLA         & Wang              & LBNE \\
 (Lang, Marchionni)                                                      & Fermilab    & Jostlein           & MicroBooNE, LBNE \\ \hline
Electronics, DAQ, and Triggering                               &    MSU      & Bromberg        & Generic R\&D, LArIAT\\
  (Bromberg, Thorn)                                                     &  BNL          &  Chen              & MicroBooNE, LBNE \\
                                                                                    &  Fermilab   &  Baller, Deptuch, Biery & MicroBooNE, LBNE \\
                                                                                    &  SMU         &  Gui                 & LBNE \\
                                                                                    &  Columbia/Nevis &  Camileri & MicroBooNE \\
                                                                                    &  Yale         &  Church             & MicroBooNE \\
                                                                                    &  Indiana Univserity & Urheim & LBNE \\
                                                                                    &  SLAC       &  Convery            & LBNE \\ \hline
Scintillation Light Detection                                         &  Indiana University & Mufson & LBNE \\                                                    
 (Katori, Mufson)                                                          &  MIT         &  Conrad              & MicroBooNE \\
                                                                                    &  CSU         &  Buchanan        & LBNE \\
                                                                                    &  LBNL        &  Gehman          & LBNE \\
                                                                                    &  Yale        &  Cavanna          & LArIAT \\
                                                                                    &  University of Chicago  &  Schmitz & LArIAT \\ \hline
Calibration and Test Beams                                        &  Fermilab  & Raaf, Rebel     & LArIAT\\ 
 (Raaf, Mauger)                                                           & Yale          & Cavanna          & LArIAT\\                                                
                                                                                   & William \& Mary & Kordosky & LArIAT\\                                                
                                                                                   & LANL        & Mauger            & CAPTAIN\\ \hline                                               
Software   (Junk, Soderberg)                                      & Fermilab   & Snider              & LArSoft\\
\hline \hline

\end{tabular}}

\end{table*}

This document has outlined several test stands that are currently operating or planned to operate in the next year.  Those test stands are listed in Table~\ref{table:operations} along with the anticipated dates of operations and primary goals.  The table also lists the experiments that benefit from those test stands.

\begin{table*}
\caption{\label{table:operations}  Test stands supporting LArTPC development in the US.  The primary contact, goals and operation duration are listed for each.  The experiments benefiting from these efforts are also listed with the collaboration spokespeople indicated as contacts.} 
\centering
\resizebox{\textwidth}{!}{\begin{tabular}{|c|c|c|c|} 

\hline \hline

 Test Stand                                & Primary Goals                                      & Operational Duration & Experiments \\
 (Contact)                                 &                                                               &                                   & Supported\\ \hline 
MTS                                         & Qualify materials for use in LArTPCs   & 2008 -                        & All\\ 
(Pordes)                                   &                                                              &                                   & \\ \hline
LAPD                                       & Demonstrate long electron lifetimes     & 2011-2013                 & All\\ 
 (Rebel)                                   &  without evacuation                               &                                   & \\ \hline
LArIAT                                      & Calibration in a charged particle beam & 2014 -                        & MicroBooNE,\\ 
(Cavanna, Kordosky,               &                                                               &                                   & LBNE \\ 
Raaf, Rebel)                            &                                                               &                                   & \\ \hline
CAPTAIN                                 & Calibration for neutrons and                 & 2014 -                        & LBNE \\ 
(Mauger)                                  & low energy neutrinos                            &                                   & \\ \hline
Tall-Bo                                     & Facility for testing                                 & 2013 -                        & All \\ 
(Pordes, Rebel)                       & photon detection techniques                &                                   & \\ \hline
CSU Photon Detection            & Facility to test LBNE                            & 2013 -                        & LBNE \\ 
(Buchanan)                              & photon detection techniques               &                                   &   \\ \hline \hline
ArgoNeuT                                & Demonstrate operation of LArTPC       & 2009-2010                 & \\ 
 (Soderberg)                            &  in neutrino beam, measure                 &                                   & \\           
                                                &  $\nu$-argon cross sections                 &                                   & \\ \hline
MicroBooNE                            & Measure $\nu$-argon cross sections,& 2014 - 2020               & \\ 
(Fleming, Zeller)                      & study MiniBooNE low energy excess  &                                  & \\ \hline                                               
LBNE                                       & Long-baseline oscillations                    & 2022 - 2032              & \\
(Diwan, Wilson)                       &                                                               &                                  & \\
\hline \hline

\end{tabular}}

\end{table*}

The currently operating test stands and experiments under construction have produced a valuable set of lessons that should be kept in mind for future development of LArTPCs.  Those lessons are high-lighted in Table~\ref{table:lessons}, which lists each lesson, the facility from which it came and a contact for more information.

\begin{table*}
\caption{\label{table:lessons}  Lessons learned from the operating test stands and experiments under construction and the section of the document in which more information may be found.} 
\centering
\resizebox{\textwidth}{!}{\begin{tabular}{|c|c|l|} 

\hline \hline

 Source        & Section                      & Lesson                                                                      \\ \hline 
MTS             & \S~\ref{sec:MTSP}    & Must measure outgassing rates for water in              \\
                     &                                   & each material to be used at temperatures $>87$~K  \\ 
                     &                                    & The temperature profile of the material                       \\
                     &                                   & determines the amount of outgassing                         \\ 
                     &                                    & The vapor flow pattern determines how                       \\
                     &                                   & much of the outgassed water reaches the liquid          \\ 
                     &                                   & The product of outgassing rate and vapor flow            \\
                     &                                    & is most important for determining electron lifetime       \\ \hline
LAPD           & \S~\ref{sec:LAPD}     & Virgin oxygen filter media must be carefully                 \\ 
                     &                                    &  regenerated to avoid self-heating                                \\ \hline
MicroBooNE & \S~\ref{sec:uBCryo}  & The cool-down system choice should be                      \\ 
                     &                                    & evaluated for potential cost savings                              \\ \hline
MicroBooNE & \S~\ref{sec:ODH}      & Oxygen Deficiency Hazard calculations should           \\ 
                     &                                   & be standardized for LArTPC projects                           \\ \hline
MicroBooNE & \S~\ref{sec:TPC}      & Construction tolerances should be examined in            \\ 
                     &                                    & light of the materials being used                                  \\ 
                     &                                   & Cleaning requirements need to be communicated       \\ 
                     &                                   & to machining shops to avoid extra effort during           \\ 
                     &                                   & construction                                                                 \\ \hline
Tall Bo          & \S~\ref{sec:scintN2}  & Nitrogen contamination must be kept to $>1$~ppm    \\
                     &                                   & in order to detect scintillation light                               \\ \hline

\hline \hline

\end{tabular}}

\end{table*}

The workshop identified two major areas of development that were high priorities in addition to the ongoing efforts.  The first is the development of stable high voltage feed throughs.  Despite the fact that ICARUS is operating reliably with a high voltage of 75~kV, few other systems have been shown to operate reliably at higher voltages.  The feed through design used by the 2~m drift distance LArTPC located in LAPD (see \S~\ref{sec:FEE},~\ref{sec:HV}) has a requirement of delivering 100~kV to the cathode, but currently can only reliably deliver 70~kV.  As described in \S~\ref{sec:HV} only a few individuals in the world have experience building high voltage feed throughs and none would say doing so is well understood.  An investment in groups desiring to contribute to making this vital component operate reliably would be well placed.  

The second major area of development that should receive attention is the operation of LArTPCs in the presence of magnetic fields.  No attempts have been made to do so at this time, however the benefits in terms of separating charged-current interactions from neutrinos and antineutrinos are obvious.  Various groups from the LBNE and $\nu$STORM experiments have expressed interest in these detectors; LBNE groups would like to use magnetized LArTPCs as a near detector while $\nu$STORM groups are interested in them for a far detector.  The first step in constructing such a detector would be a small LArTPC operating inside a magnet such as those available at Fermilab in the test beam facility.

\section{Acknowledgments}
The speakers at this workshop represented many colleagues from both collaborations and local university groups.  The work they presented was supported by a variety of funding agencies.  The following statements of acknowledgement were supplied by the speakers and have not been edited.

We thank the ICARUS collaboration for sharing their experiences and lessons learned in building and operating the first large LArTPC.  The High Energy Astrophysics Group at Indiana University is supported by the U.S. Department of Energy Office of Science with grant DE-FG02-91ER40661 to Indiana University and LBNE project funding from Brookhaven National Laboratory with grant BNL 240296 to Indiana University.  The LArIAT collaboration is supported by the U.S. Department of Energy Office of Science and the National Science Foundation.  The CAPTAIN detector has been designed and is being built by the Physics and the Theory divisions of Los Alamos National Laboratory under the auspices of the LDRD program.  T. Strauss spoke on behalf of the Albert Einstein Center, Laboratory of High Energy Physics of the University of Bern.  The MicroBooNE and LBNE collaborations have participated in the development of cold electronics as supported by the U.S. Department of Energy Office of Science.

\bibliographystyle{JHEP}
\bibliography{sumbib}

\end{document}